\documentclass[amsmath,twocolumn,prl]{revtex4}
\usepackage{amssymb}
\usepackage{amsbsy}
\usepackage{amsmath}
\usepackage{graphicx}
\usepackage{dcolumn}
\usepackage{bm}
\usepackage{natbib}
\usepackage{bbm}

\usepackage{hyperref}
\usepackage[usenames,dvipsnames]{xcolor}
\definecolor{med-blue}{RGB}{25,25,112}
\hypersetup{colorlinks, linkcolor={red},citecolor={blue}, urlcolor={MidnightBlue}}

\def\be{\begin{equation}}
\def\ee{\end{equation}}

\begin{document}
\title{Freezing a Quantum Magnet by Repeated Quantum Interference:\\ An Experimental Realization}
\author{Swathi S. Hegde}
\altaffiliation{These authors contributed equally to this work.}
\author{Hemant Katiyar}
\altaffiliation{These authors contributed equally to this work.}
\author{T. S. Mahesh}
\email{mahesh.ts@iiserpune.ac.in}
\affiliation{Department of Physics and NMR Research Center, 
Indian Institute of Science Education and Research, Pune 411008, India}
\author{Arnab Das}
\email{arnab.das.physics@gmail.com}
\affiliation{Theoretical Physics Department, Indian Association for the Cultivation of Science, Kolkata 700032, India}
\affiliation{Max-Planck-Institut f\"{u}r Physik komplexer Systeme, 01187 Dresden, Germany}

\begin{abstract} 
We experimentally demonstrate the phenomenon of dynamical many-body freezing
in a periodically driven Ising chain. Theoretically [Phys. Rev. B {\bf 82}, 172402 (2010)], 
for certain values of the drive parameters 
{\it all fundamental degrees of freedom} contributing to the response dynamics 
freeze for {\it all time} and for {\it arbitrary initial state}. 
Also, since the condition of freezing involves only the drive parameters and
not on the quantization of the momentum (i.e., the system-size), 
our simulation with a small (3-spin) chain captures all salient features of 
the freezing phenomenon predicted for the infinite chain. 
Using optimal control techniques, we realize high-fidelity cosine modulated drive,
and observe non-monotonic freezing of magnetization at specific frequencies of modulation.
Time-evolution of the excitations in momentum space has been tracked directly 
through magnetization measurements.
\end{abstract}

%

\maketitle

\section{Introduction}
The study of non-equilibrium dynamics has emerged as one of the central 
topics in quantum many-body physics,
thanks to the recent advancements in various experimental techniques 
\cite{NMR-1,NMR-2,NMR-3, NMR-4, NMR-5, NMR-6,NMR-7, Bloch-1, Bloch-2, Blatt}. 
In particular, the paradigm of periodically driven 
many-body systems has hosted quite a few interesting theoretical (see, e.g., 
\cite{Andre-1,Andre-2,AAR-PRL,AAR-Generic,Arimondo-Rev,Gauge-1a,Gauge-1b,Gauge-2,Prosen-1,Kris-Periodic,Bastidas,victor,Sthitadhi}) 
and experimental results (see, e.g., \cite{Arimondo-1,Arimondo-2,Bloch-Periodic,Shengstock,haeberlen,cavanagh1995protein}) 
in recent days. We add to this list, experimental verification of a non-equilibrium 
freezing phenomenon, where quantum interference brings about a surprising 
counter-classical scenario \cite{AD-DQH}.
 
In general, response of a simple driven system may ``freeze" 
if the drive rate is much faster compared to the characteristic relaxation rate of the system.
The mechanism is intuitively simple -- the system does not get sufficient time 
to change itself significantly within the duration of the drive. 
Thus, faster the drive is, more frozen is the response. 
This intuition is manifested in many 
celebrated results in both classical and quantum physics, 
like Kibble-Zurek scaling law for defect generation in classical \cite{Kibble,Zurek} 
and quantum \cite{Bogdan-LZ,Zurek-Dorner,Dziarmaga} phase transitions,
Landau-Zener excitation probability \cite{Landau,Zener}, phenomenon of 
classical dynamical hysteresis \cite{BKC-RMP} to name a few. It forms the basis of the so called
sudden approximation in quantum mechanics \cite{Messiah}, and its more elaborate descendent 
- adiabatic-impulse approximation which applies to classical
and quantum systems alike \cite{Zurek,DZ-LZ}. 
For a periodically driven system, this intuition implies a stronger freezing for a higher 
frequency of the drive. 

However, in certain cases this intuition may grossly fail due to quantum interference of excitation
amplitudes. In our experiment we demonstrate one such phenomenon, dubbed as dynamical many-body freezing 
\cite{AD-DQH,AD-SDG,AD-RM,Anatoli-Periodic,Kris-Periodic,BKC-Book}, where repeated quantum interference results in 
{\it strongly non-monotonic} freezing behaviour with respect to the driving rate
in a periodically driven quantum magnet, contrasting the above intuitive scenario.
Most surprisingly, {\it absolute freezing}  (i.e. freezing of 
all excitations contributing to the response dynamics) is observed for certain particular values of drive parameters 
in a large class of integrable models for various kinds of periodic drive \cite{AD-DQH,AD-SDG,AD-RM,Anatoli-Periodic,BKC-Book}.

This phenomenon is akin to the well-known single-particle phenomena of dynamical localization~\cite{Dunlap},
and coherent destruction of tunneling~\cite{CDT}, where a quantum particle is kept localized in space 
under an externally applied periodic drive. But it is not only a many-body extension of dynamical localization,
it is also a much more drastic version of it  -- here the freezing is not limited to initial states localized in real space
as happens in dynamical localization, but applies for {\it any} initial state in the Hilbert space \cite{AD-DQH}.


Major experimental work has been done in recent years (see, e.g., \cite{Ref2,Ref1}) 
in probing equilibrium quantum critical properties, e.g., locating a quantum critical point of an Ising 
magnet using NMR. Here we take a major step forward by experimentally demonstrating a non-equilibrium 
phenomena in a  {\it coherently driven many-body system}, where quantum coherence plays a surprising 
role as we discuss here.

  
The rest of the paper is organized as follows. First, we outline the theory and the phenomenology  
of the dynamical many-body freezing, particularly for our finite-size experimental system. Then we
discuss the experimental setup and realization of the phenomenon followed by the result and its analysis,
and finally we conclude with an outlook.  

\section{Dynamical Many-body Freezing in Ising Chain: Theory} 

\subsection{Review of the Phenomenon in an Infinite Chain}
First we review the phenomenon in an infinite one-dimensional Ising ring 
that is being subjected to a transverse periodic-driving field of 
amplitude $h_0$, frequency $\omega$ (period $\tau = 2\pi/\omega$), and
represented by the Hamiltonian,
$
{\cal H}(t) = -\frac{1}{2} [{\cal J} \sum_{i=1}^{\infty} \sigma_{i}^{z}\sigma_{i+1}^{z} 
+ h_{0}\cos{(\omega t)}\sum_{i=1}^{\infty}\sigma_{i}^{x}].
$
Here ${\cal J}$ is the strength of the nearest neighbour Ising interaction
and $\sigma_{i}^{x/z}$ are Pauli matrices.
A dynamical freezing parameter $Q$, defined as the long-time average of magnetization 
in $ x $-direction, $m^{x}(t)$, is used for quantifying the degree of freezing 
via the anomalous DC response. 
\begin{equation}
Q = \lim_{{\cal T} \to \infty}\frac{1}{\cal T}\int\limits_{0}^{\cal T} m^x(t) dt,
\label{Q_def}
\end{equation} 
where ${\cal T}$ is the duration of total evolution.
\noindent
Suppose that at $t=0$ the system is at a state completely polarized in
the $x$-direction. Then the adiabatic limit ($\omega/L \to 0$, where $ L $ is the system-size) 
corresponds to $Q = 0$ in $L \to \infty$ limit
(since the average of the transverse field is zero over each cycle), and the impulse or ``freezing" limit of
($\omega \to \infty$) corresponds to $Q=1/2$. Classical intuition of ``faster drive means lesser time to react" 
would suggest a monotonic behavior of $Q$ as a function of $\omega$ interpolating between these two regimes. 
But quantum interference leads to a strongly non-monotonic freezing behavior for $\omega \gg {\cal J}.$  
In the limit $L\to\infty$, $Q$ is accurately given by a closed form analytical formula \cite{AD-DQH} 
\begin{equation}
Q = \frac{1}{1 + |J_{0}(2h_{0}/\omega)|},
\label{Q_Ana}
\end{equation}    
\noindent 
where $J_{0}(.)$ denotes the ordinary Bessel function of order $0$. 

The gist of the result in Eq.~(\ref{Q_Ana}) is summarized below.
First, the average magnetization $Q$, which quantifies the degree of freezing, is a 
highly non-monotonic function of $\omega$ in contrast to what is expected intuitively.
Second, there are certain values of the ratio of the amplitude $h_{0}$ and frequency $\omega$ of the
driving, at which
\begin{equation} 
\label{fc}
J_{0}(2h_{0}/\omega)=0, 
\end{equation}
we have $Q=1$, which means the time-average of $m^{x}(t)$
is equal to the initial $m^{x}$, which was set to the maximum value (unity) which $m^{x}$ can assume.
This is possible if $m^x$ has remained {\it absolutely frozen for all time}. Analysis shows this 
happens because population dynamics of each fundamental degree of freedom (which happens to be 
independent fermionic excitations in momentum space) freezes absolutely \cite{AD-DQH}. 

\begin{figure}[b]
\centering
\includegraphics[width=\linewidth]{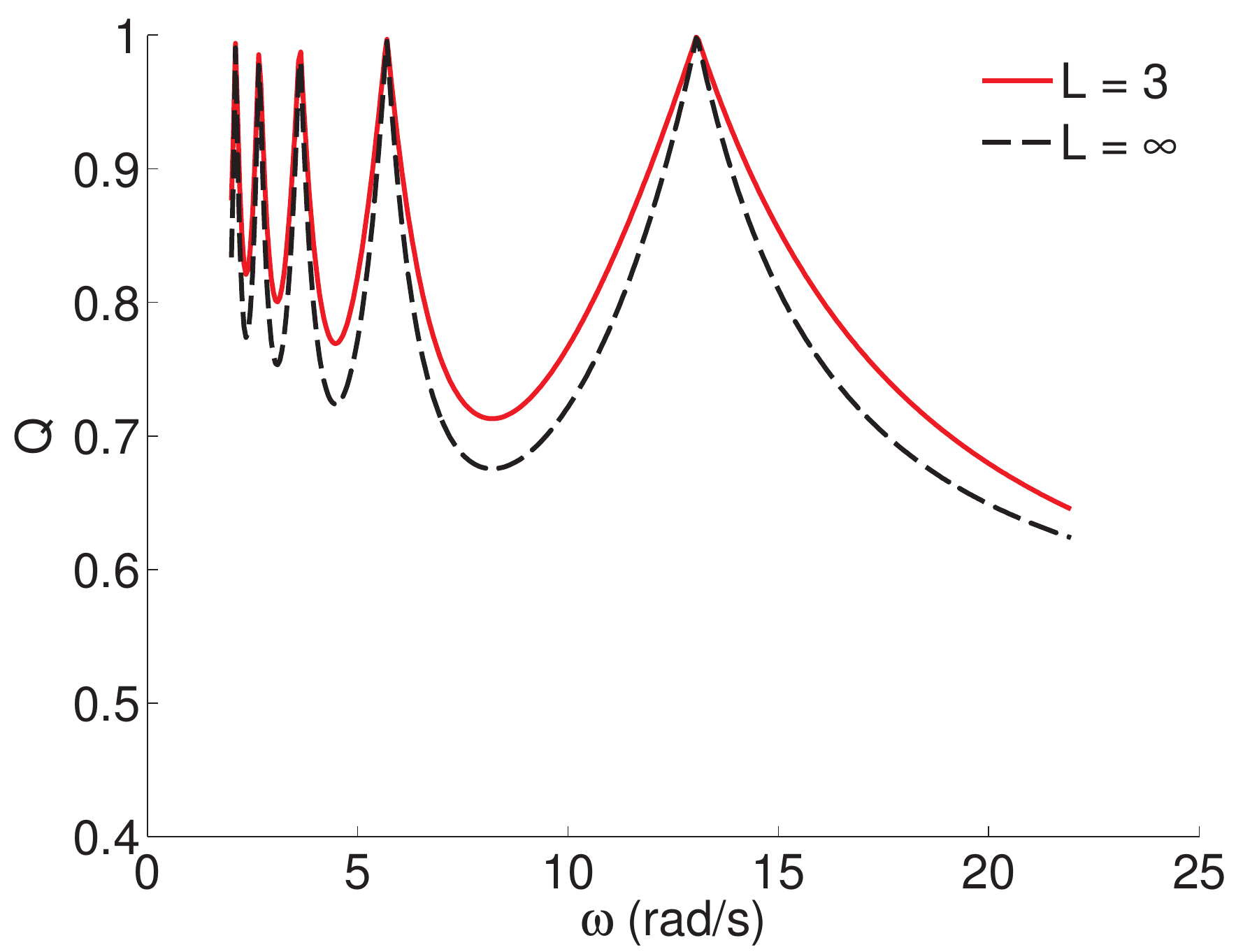}
\caption{{\small Comparison of analytical results for $Q$ (obtained for the $\omega \gg {\mathcal J}$ limit 
for $L\to\infty$ (Eq.~\ref{Q_Ana}) and $L=3$ (Eq.~\ref{Q-L3}) for $h_{0}=5\pi, {\mathcal J} = h_0/20$ (in rad/s). 
Strong non-monotonic freezing behavior with respect to the drive frequency $\omega$ is visible,
in contrast to the intuition of faster drive $\Rightarrow$ lesser time to respond $\Rightarrow$
more frozen response. Almost absolute freezing $Q\approx 1$ is visible for certain special 
values of $\omega,h_{0}$ satisfying the freezing condition (Eq~\ref{fc}). 
Our experiment captures all these features remarkably well 
(Fig.~\ref{qplot}).
}}
\label{Q-L-3-Infty-Ana}
\end{figure}

\subsection{Dynamical Many-body Freezing for $L=3$}
In this subsection we discuss the fate of dynamical freezing for a chain of size $L=3,$ which is the size of our
experimentally realized sample. Analysis of this particular case is important not only for matching
with the experiment, but it has an important finite-size effect which vanishes in the $L\to\infty$
limit. Nevertheless, we show that $L=3$ results captures all the essential features of the infinite 
size limit. Moreover, this analysis allows us to illustrate the pivotal role quantum interference
plays in the phenomenon.    

The Hamiltonian we consider is
\be
{\cal H} = -\frac{1}{2}\left[{\cal J}\sum_{i}^{3}\sigma_{i}^{z}\sigma_{i+1}^{z} + h_{0}\cos{(\omega t)}\sum_{i}^{3}\sigma^{x}_{i} \right],
\label{H-Spin}
\ee
\noindent with Periodic Boundary Condition $\sigma_{4} = \sigma_{1}$.
The eigen-problem of the above Hamiltonian can be solved 
analytically by Jordan-Wigner transformation followed by 
Fourier transform for any $L$ (see. e.g., \cite{LSM,Mattis,BKC-Book}). 
The above Hamiltonian can be mapped to independent fermionic Hamiltonians
in momentum space such that only fermions with equal but opposite momenta $\pm k$ are coupled:
${\cal H}(t) = \prod_{k>0} H_{k}$, with
\begin{equation}
 H_{k} = -E_{k}(c_{k}^{\dagger}c_{k}  -c_{-k}c_{-k}^{\dagger}) + 
i\Delta_{k}(c_{-k}c_{k} - c_{k}^{\dagger}c_{-k}^{\dagger}), 
\label{HXYk}
\end{equation}
\noindent where
$E_{k} =  h_{0}\cos{\omega t} + {\cal J}\cos{k}$, $\Delta_{k} = {\cal J}\sin{k}$, and $h_{x}(t) = h_{0}\cos{\omega t}$.
The quantization of $k$ depends on the parity of the fermion number (i.e., whether it is odd/even) which is
conserved throughout the dynamics (though the parity in the ground state depends on the sign of $h_{x}$, 
which means the ground state can never be followed adiabatically - see \cite{Marek}. From dynamical 
point of view, this happens because one of the three modes remains absolutely frozen irrespective of how
slow the drive is, as discussed below).
Here for definiteness, we start with the ground-state of $\cal{H}$ for any parameter value, which will always have
an even number of fermions. This leads to the quantization 
$k = -\pi, \pm\pi/3$ (given our choice of our Brillouin zone). Now we note that $k = -\pi$ has no
partner to pair with, while the only pair formed is for $k = \pm \pi/3$. Thus we can work with a
Hilbert space spanned by the vectors 
$\{[|0_{-\pi/3},0_{\pi/3}\rangle, |1_{-\pi/3},1_{\pi/3}\rangle ]\otimes[|0_{-\pi}\rangle,|1_{-\pi}\rangle]\}$.
In this representation the initial state fully polarized in $+x$-direction reads 
$|\psi(0)\rangle = |1_{-\pi/3},1_{\pi/3}\rangle\otimes |1_{-\pi}\rangle$. 
Henceforth we denote $|0_{-\pi/3},0_{\pi/3}\rangle$ by $|0_{\pi/3}\rangle$ and 
$|1_{-\pi/3},1_{\pi/3}\rangle$ by $|1_{\pi/3}\rangle.$
Noting that the $k = -\pi$
mode has no dynamics, the state at any time can be given by
\begin{equation}
|\psi(t)\rangle = [u_{\pi/3}|0_{\pi/3}\rangle + v_{\pi/3}|1_{\pi/3}\rangle]\otimes|1_{-\pi}\rangle,
\label{psi_t}
\end{equation}
\noindent The time-dependent transverse magnetization 
in this notation 
reads (for the initial state fully polarized in $+x$-direction),
\be
m^{x}(t) = \frac{2}{3} \sum_{k=-\pi,\pm\pi/3}\left\langle c_{k}^{\dagger}c_{k} \right\rangle - 1 = \frac{4}{3}|v_{\pi/3}(t)|^{2} - \frac{1}{3}
\label{mx-3}
\ee    
Now $\{u_{\pi/3}, v_{\pi/3}\}$ (with a suitably adjusted common phase factor; 
                               see~\footnote{To get this form, 
                               we have added a term $(\delta\cos{k}){\mathcal I}_{k}$ to the Hamiltonian,
                               where ${\mathcal I}_{k}$ is the $2\times 2$ identity matrix in the representation of Eq.~(\ref{H2}),
                               which contributes nothing but an additional overall phase to the wavefunction.}) 
satisfies following time-dependent Schr\"{o}dinger equation:
\begin{equation}
i\frac{\partial}{\partial t}
\begin{bmatrix}
u_{\pi/3}(t) \\
v_{\pi/3}(t)
\end{bmatrix} =
\begin{bmatrix}
E_{\pi/3} & i\Delta_{\pi/3}\\
-i\Delta_{\pi/3} & -E_{\pi/3}
\end{bmatrix}
\begin{bmatrix}
u_{\pi/3}(t) \\
v_{\pi/3}(t)
\end{bmatrix},
\label{H2}
\end{equation}
\noindent
where $E_{\pi/3} = h_{0}\cos(\omega t) + {\cal J}\cos{(\pi/3)}$ and $\Delta_{\pi/3} = {\cal J}\sin{(\pi/3)}.$
Exact solution of the above $2\times 2$ matrix equation 
is not know but analytical solution can be obtained 
under a rotating wave approximation in the fast driving regime ($\omega \gg 2{\mathcal J}$) \cite{AD-DQH}.
For the initial condition mentioned above, the solution for $|v_{\pi/3}|^2$ reads
\begin{eqnarray} \nonumber
\label{phik}
|v_{\pi/3}|^2 &=& 1 - A_{\pi/3}^{2}\sin^{2}{(\phi_{\pi/3}t)}, ~~{\rm where} \\ 
\phi_{\pi/3} &=& |{\mathcal J}|\sqrt{J_{0}^{2}(2h_{0}/\omega)\sin^{2}{\left(\frac{\pi}{3}\right)} 
+ \cos^{2}{\left(\frac{\pi}{3}\right)}}, \\
\label{A-L3}
A_{\pi/3} &=& \frac{J_{0}(2h_{0}/\omega){\mathcal J}\sin{\left(\frac{\pi}{3}\right)}}{\phi_{\pi/3}}
\end{eqnarray} 
%
\noindent
This gives
\begin{eqnarray}
\label{mx-3}
m^x(t) &=& 
Q (L=3) + \frac{2A_{\pi/3}^{2}}{3}\cos{(2\phi_{\pi/3}t)}, \\
\label{Q-L3}
Q (L=3) &=& 1 - \frac{2}{3}A_{\pi/3}^{2} = \frac{1+|J_{0}(2h_{0}/\omega)|}{1+3|J_{0}(2h_{0}/\omega)|},
\end{eqnarray}
\noindent Comparing this with Eq~(\ref{Q_Ana}) we see that $Q$ has maximum value (unity) in both cases
of $L=3$ and $L\to\infty$ for $J_{0}(2h_{0}/\omega)=0$. Also, since $|J_{0}(2h_{0}/\omega)| \le 1$,
$Q(L=3) \ge Q(L\to\infty).$ Our experimental results accurately capture these features (Fig.~\ref{qplot}).

\begin{figure}[b]
\centering
\includegraphics[width=7.7cm]{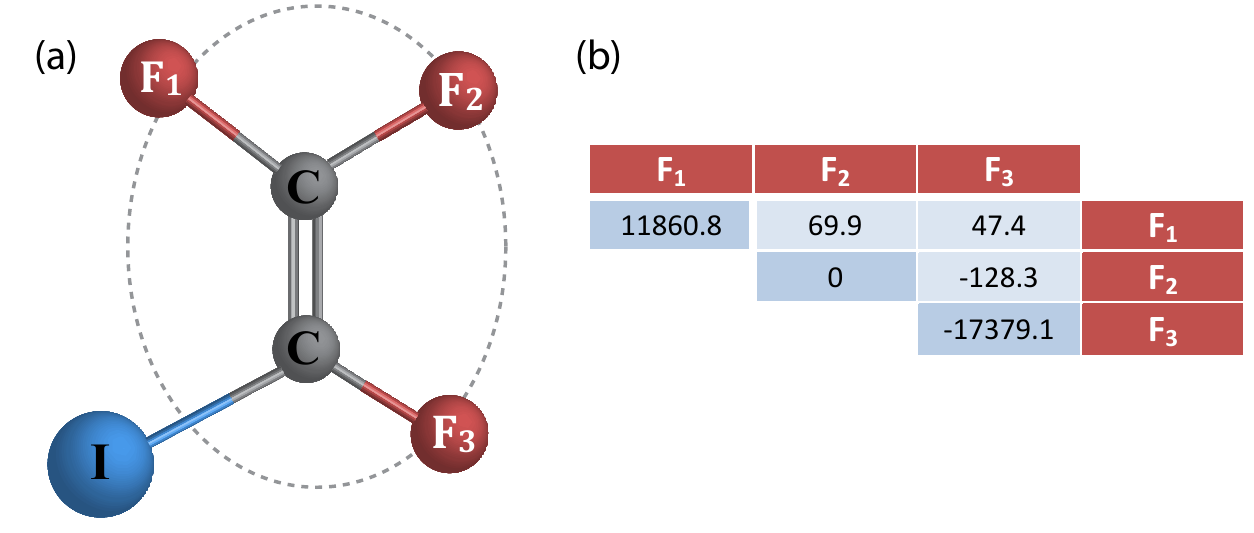}
\caption{{\small (Color online)
(a) Molecular structure and (b) chemical 
shifts (diagonal terms) and 
 $J$-coupling (off-diagonal terms)
 in Hz of trifluoroiodoethylene.
}}
\label{molcule}
\end{figure}
\section{Experimental Realization}
We simulate the above freezing phenomenon in the 3-spin chain given by Eq. 
(\ref{H-Spin}). Going beyond the
theoretically explored cases of pure initial states, we also demonstrate 
strong freezing for a mixed initial state. 
We use the three spin-1/2 $^{19}$F nuclei
of trifluoroiodoethylene dissolved in acetone-D6 as our 3-spin Ising system.
The molecular structure of the system is shown in Fig.~\ref{molcule}(a).
All the experiments described below were carried out using a Bruker 500 MHz
NMR spectrometer at an ambient temperature of 290 K. 
The simulation involves preparing a certain initial state, evolving it under the Hamiltonian 
${\cal H}(t)$ for a total time $T$, while
measuring the response of the system in intervals of time $ \tau $.

The thermal equilibrium state for the NMR spin-system is $(\mathrm{I}/8+\epsilon \sum_{i=1}^3 \sigma_i^z/2)$ where $\mathrm{I}$ represents uniform background population, which does not evolve under unitary transformations and henceforth ignored. Here $\epsilon$ is the polarization of the spin (see \cite{suppl}). 
The response of the system measured as transverse magnetization is given by
\begin{equation}
m^{x}(t) = \frac{1}{m_0}\mathrm{Tr}\left[\rho(t)\left(\sum_{i=1}^3\sigma_i^x/2\right)\right].
\label{mxt}
\end{equation} 
Here $\rho(t)$ is the instantaneous
density matrix of the driven system and $ m_0$
is a normalization factor \cite{suppl} given by the maximum possible transverse magnetization \textit{i.e.}
$ m_0 =\mathrm{Tr}\left[ (\sum_{i=1}^3 \sigma_i^x/2)(\sum_{i=1}^3 \sigma_i^x/2) \right]$ .
We performed a set of experiments for each of the following two different initial states.  
The first initial state is $\rho(0) = \sum_{i=1}^3 \sigma_i^x/2$ (i.e. $ m^x(0)=1 $, 
fully polarized in $+x-$direction) obtained by applying a global $R_y(\pi/2) = \exp\{-i (\pi/2) \sum_i^3 \sigma^y_i/2\}$ rotation
on the thermal state. This state is approximately the 
ground state of our Hamiltonian (\ref{H-Spin}) at $t=0$, if $h_{0} \gg {\cal J}$.
The second initial state is
$\rho(0) = \sum_{i=1}^3 (\sigma_i^x/2 + \sigma_i^z\sqrt{3}/2)/2$ (i.e. $ m^x(0)=0.5 $) 
obtained by applying a global $R_y(\pi/6)$ rotation on the thermal state. 
Note that this state is not the ground state of Hamiltonian (\ref{H-Spin}).
For convenience, we refer these two sets of experiments by their initial magnetization values, $ m^x(0) $.

The internal Hamiltonian for the NMR system is given by
\begin{eqnarray}
{\cal H}_{\mathrm{int}} = -\pi \sum\limits_{i=1}^{3}\nu_i \sigma_i^z 
+ \frac{\pi}{2} \sum\limits_{\substack{i,j=1 \\ i<j}}^{3} J_{ij} \sigma_i^z\sigma_j^z,
\label{hint}
\end{eqnarray}
where the first term represents Zeeman part and the second term represents spin-spin interaction part.
Here $\nu_i$ is the resonance frequency of the $i$th spin in the
rotating frame, and $J_{ij}$ is the strength of the
indirect spin-spin interaction between spins $i$ and $j$.  
The parameters of internal Hamiltonian for the above spin-system
are shown in Fig.~\ref{molcule} (b).

In order to simulate the time-dependent Hamiltonian ${\cal H}(t)$
using ${\cal H}_{\mathrm{int}}$,
we need to (i) cancel the evolution under Zeeman interaction,
(ii) bring out an effective Ising interaction of strength ${\cal J}$,
and (iii) add the oscillatory drive $(-h_0/2) \cos{(\omega t)}$ along the x-direction.
Equivalently, the evolution under the Hamiltonian ${\cal{H}}(t)$ in Eq. (\ref{H-Spin}) is realized by 
designing an RF modulation using a standard optimal control technique 
such that the effective evolution for a small time step  $\delta t$ is
as close to $\exp(-i {\cal H }\delta t)$ as possible \cite{Khaneja}.
The procedure involves discretizing time-duration $T$ into
$M$ equal steps each of duration $\delta t = T/M$ such that  
the propagators for the discretized Hamiltonian are
\begin{eqnarray}
U_k = \exp{[-i \delta t {\cal H}(k\delta t)]},
\label{uk}
\end{eqnarray}
for $k \in \{1,2,\cdots,M\}$.
The overall evolution for a time $k\delta t$ is given by
the unitary
$U(k \delta t) = U_k U_{k-1} \cdots U_2 U_1$.
\begin{figure}[b]
\centering
\hspace*{-.9cm}
\includegraphics[width=1.09\linewidth]{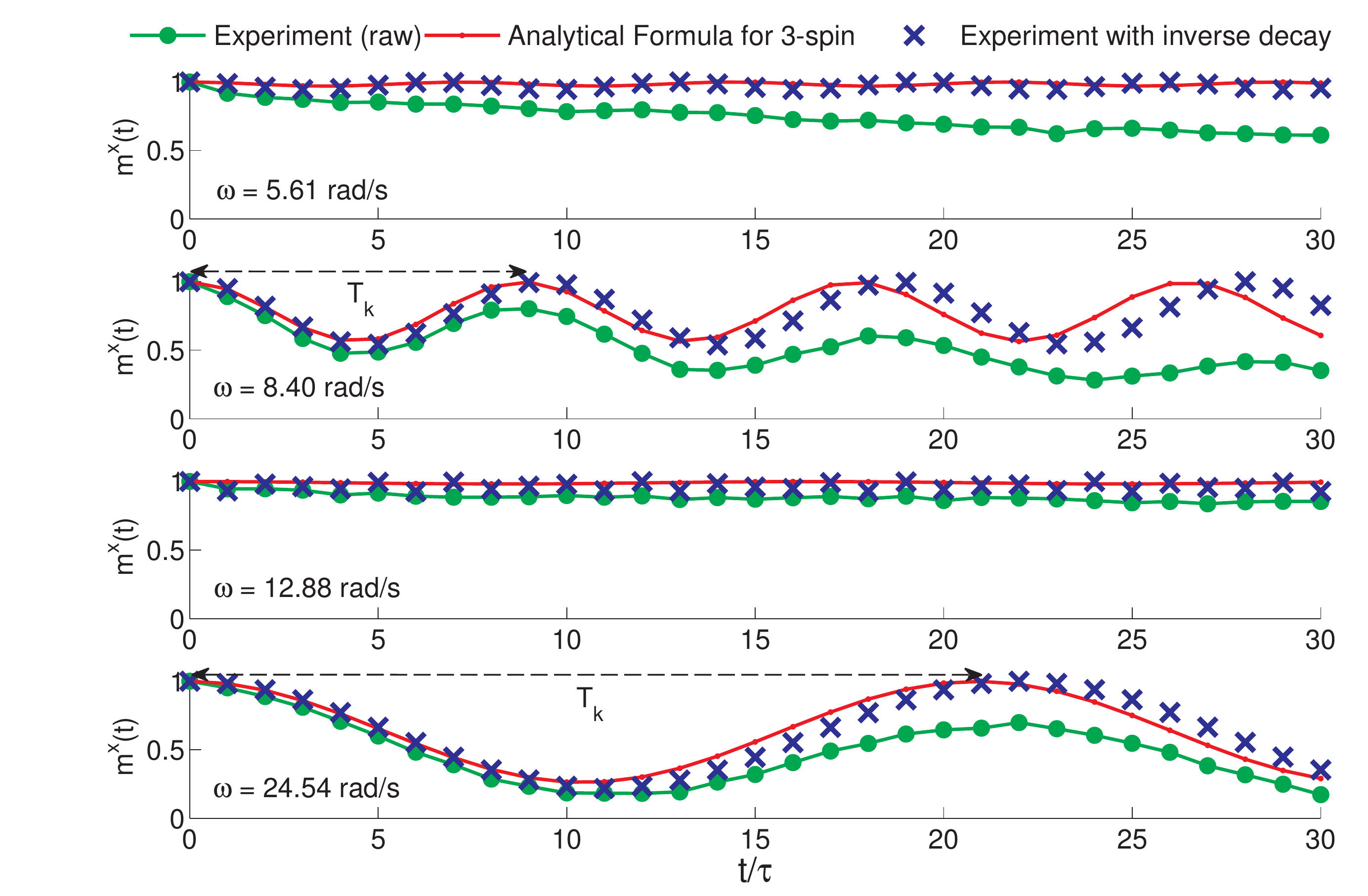}
\caption{{\small (Color online)
(Color online) Evolution of $m^x$ with time 
for $m^x(0) = 1$. The period ($\tau$) of oscillation of $m^{x}$  
visible  from the experimental data (marked for $\omega = 8.40$, $24.54$ rad/s) 
corresponds to the frequency $2\phi_{k}$ (period $T_{k} = 2\pi/\phi_{k}$) 
of population oscillation for the 
fundamental excitation in momentum space for $k=\pi/3$ 
given in Eq.~(\ref{phik}). The analytical curves represent Eq.~(\ref{mx-3}).
%
}}
\label{result1}
\end{figure}
The response $ m^x(t) $ (refer Eq.~(\ref{mxt})), is measured 
at time instants $t=n\tau$, for $n=0,1,2,\cdots,N$, where $N = T/\tau$ 
corresponds to the last measurement.  Therefore we only need to
generate the propagators $U(n\tau)$.
We used Gradient Ascent Pulse Engineering (GRAPE) method
to design amplitude and phase modulated radio frequency (RF) pulses
which effectively realize these propagators \cite{Khaneja}.
\begin{figure*}
\hspace*{-.4cm}
\includegraphics[width=0.8\linewidth]{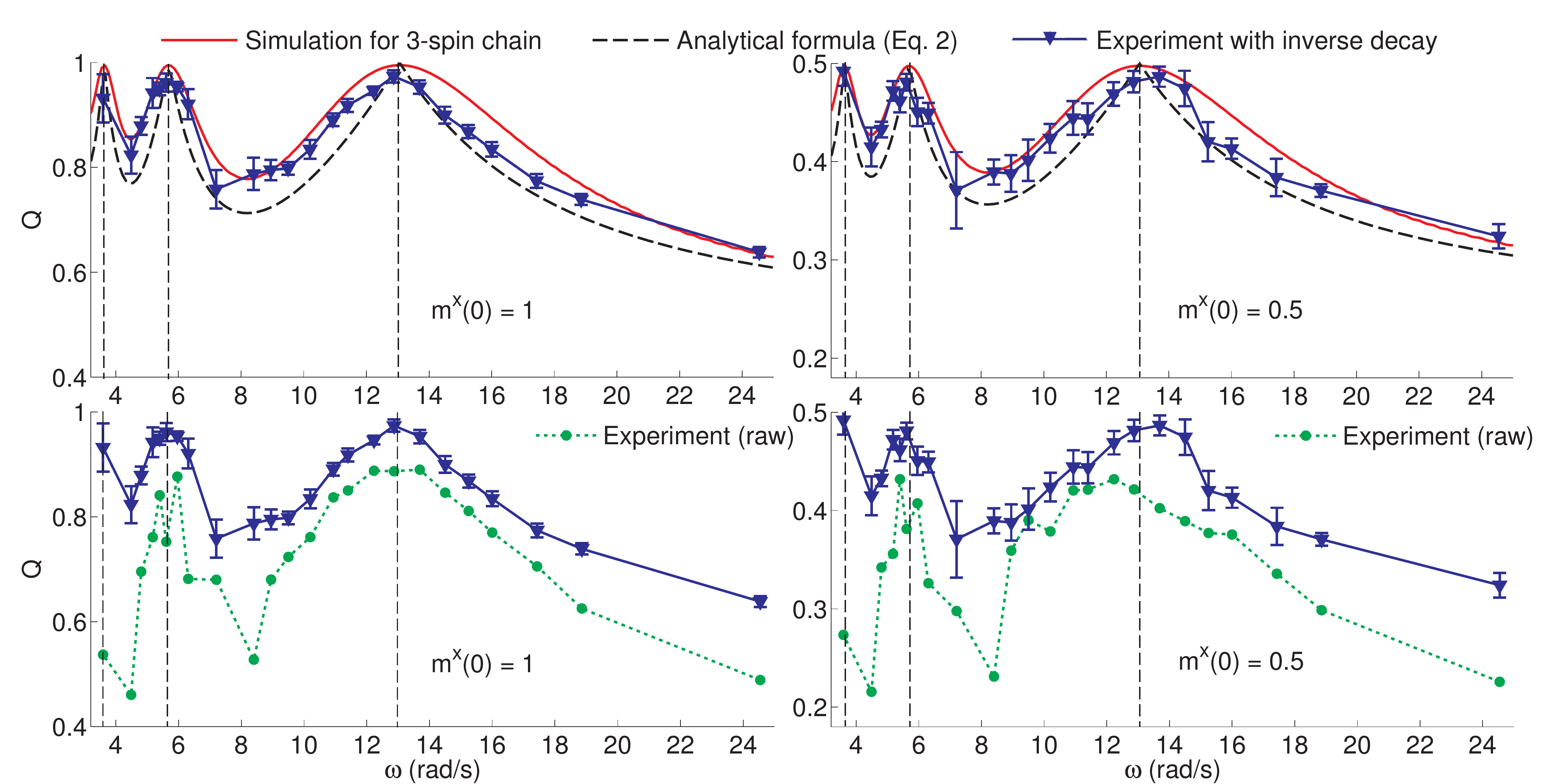}
\caption{(Color online) {\small 
$Q$ vs $\omega$ for different initial conditions:
{ Left panels} for the initial state corresponding to $m^{x}(0) = 1$, { Right panels}
for the initial state corresponding to $m^{x}(0) = 0.5$.
{ Top frames} compare the final experimental data (with
inverse decay correction) for 3-spins with the theoretical
results for $L=3$ and $L \to \infty$. Simulation for L=3 is done by solving the exact time-dependent Scr\"{o}dinger equation numerically. $Q$ is obtained by measuring $m^{x}$ after each cycle and carrying out the stroboscopic average over 30 cycles (as done in the experiment). The vertical lines indicate
the position of the freezing peaks.
The { bottom frames} compare raw experimental data with the decay-corrected data. 
Both the data captures all the essential features of that predicted
theoretically for $L\to\infty.$ 
}}
\label{qplot}
\end{figure*}

The dynamical freezing parameter for this discretized set of measurements
is defined as
$Q = \frac{1}{N+1}\sum\limits_{n=0}^{N}m^x(n\tau).$
The experiments were carried out for 24 
different values of $\omega$ ranging from 3.59 to 24.54 rad/s.
In each case we chose the measurement interval as $\tau=2\pi/\omega$, the period of the driving
frequency. We discretized each cycle of the periodic Hamiltonian ${\cal H}(t)$ into 11 steps
and constructed the propagators $U(n \tau)$  for $N=30$ cycles, i.e., 
$n = 0,1,2,\cdots, N$.  Here maximum number of cycles was limited by
the duty-cycle of the RF channel. 
The one-cycle propagator
$U(\tau) = U_{11}U_{10}\cdots U_2 U_1$, where $U_k$s are given by 
Eq.~(\ref{uk}). Since ${\cal H}(t)$ has a period of $\tau$, $U(n\tau) = U^n(\tau)$.
All numerically generated GRAPE pulses were optimized against
RF inhomogeneity and had an average Hilbert-Schmidt fidelity greater than or equal to $ 0.99 $.
For the 30th cycle, the overall duration of the GRAPE pulses were
ranging between 48 ms to 300 ms for different $\omega$ values.

The overall experimental sequence can be represented by the unitary
$U(n\tau) \cdot R_y(\pi/2)$ for $ m^x(0)=1 $ and $U(n\tau)\cdot R_y(\pi/6)$ for $ m^x(0)=0.5$. 
 The net-transverse magnetization corresponding 
to the observable $\sum_{i=1}^3\sigma_i^x/2$ was measured as the intensity of 
the real part of the NMR signal acquired in a quadrature mode and is
normalized by the reference signal obtained with only  $R_y(\pi/2)$ pulse \cite{suppl}.

\begin{figure}[h]
\centering
\hspace*{-0.6cm}
\includegraphics[width=9.5cm]{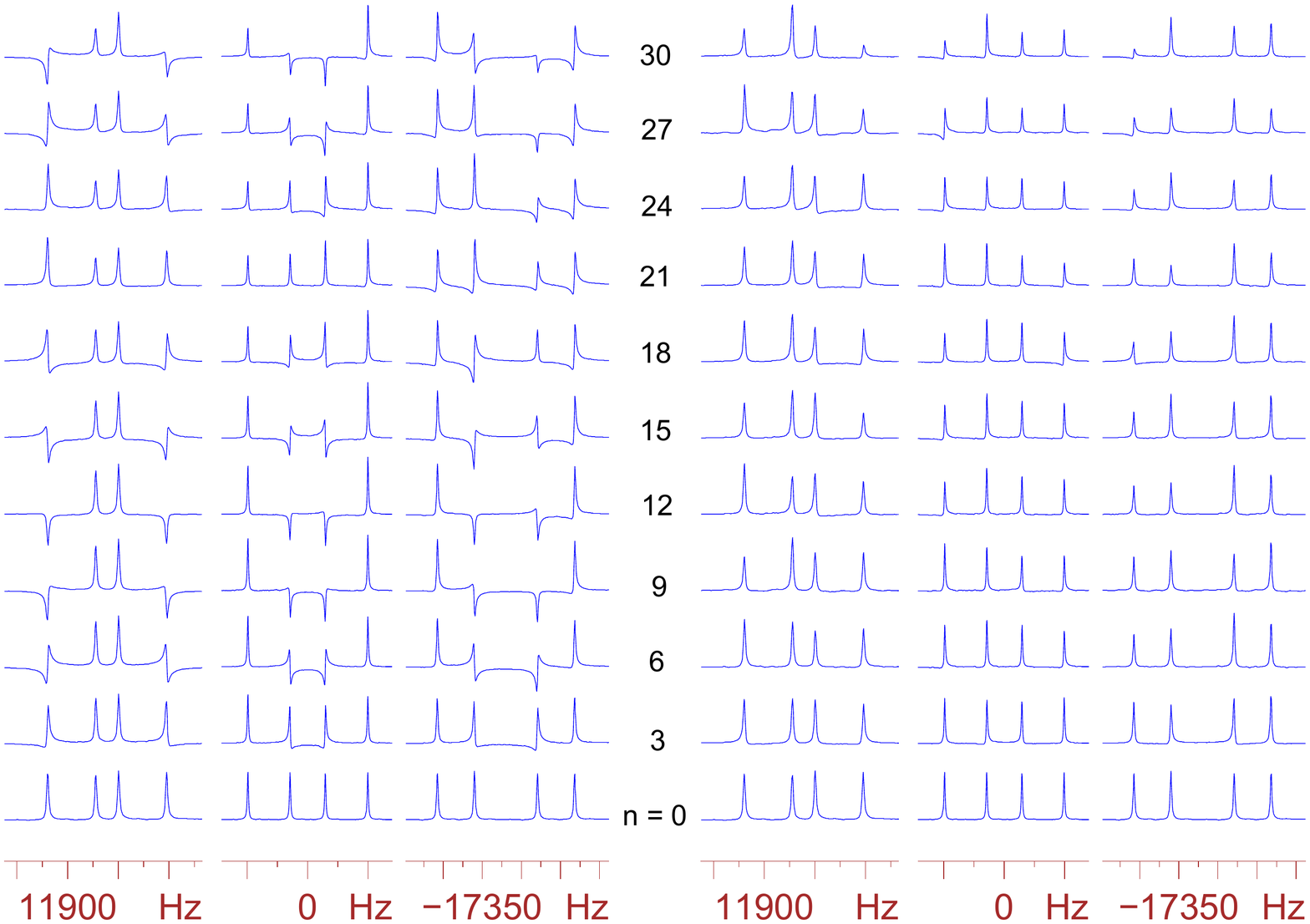}
\caption{{\small (Color online) 
$^{19}$F spectra corresponding to a non-freezing condition
(left column; $\omega=24.54$ rad/s) and a freezing condition
(right column; $\omega = 5.61$ rad/s) for $ m^x(0)=1 $.  Only spectra for
the indicated time instants ($n = t/\tau$) are shown. In contrast
to the classical intuition, the response is clearly seen to be
much strongly frozen for the much lower driving frequency 
($\omega = 5.61$ rad/s) due to quantum interference, as predicted by the theory. 
}}
\label{spec}
\end{figure}

\noindent
{\it Results:}
The main result of our experiment is summarized in Figs. \ref{result1}, \ref{qplot} and \ref{spec}.
In Fig.~\ref{result1} dynamics of $m^{x}$ as the function of time is shown for different values of $\omega$,
and for parameters $h_0 = 5\pi$ and ${\cal J} = h_0/20$ (in rad/s).
The freezing of the dynamics is clearly visible close to the theoretically predicted values of $\omega$
($\approx$ 5.61 and 12.88 rad/sec, corresponding to $J_{0}(2h_{0}/\omega) \approx 0$). 
The oscillation of $m^x(t)$ with time allows us to directly read off 
the frequency of the population oscillation
of the underlying fermionic mode $k = \pi/3$ in the momentum space using Eq.~(\ref{mx-3}).
However, from the raw experimental data (green line-filled-circle), we see $m^x$ decays steadily even
under the maximal freezing conditions. This decay is due to decoherence, transverse 
relaxation ($T_{1\rho}$) \cite{kowalewski2006nuclear} as well as decay caused by
the spatial inhomogeneities in RF amplitudes \cite{cavanagh1995protein}.  
For all other values of $\omega$, within our chosen $\omega$ range, we see
oscillations in $x$-magnetization in addition to the decay. 
In order to confirm this, we took into account the effect of decay by
fitting the experimental data points $m^x(t)$ with the standard 
decay functions, 
i.e., $m^x(t) = \alpha+[\beta+\gamma \cos(ct)]e^{-t/T_d}$ where $\alpha$, $\beta$, $\gamma$, $c$ and $T_d$ are the fitting parameters.  
Here $T_{d}$ is the decay constant and is obtained by the fitting $m^x(t)$ to the above function \cite{suppl}.
The resulting data points are shown in Fig.~\ref{result1} 
as crosses. 

Fig.~\ref{qplot} shows $Q$-vs-$\omega$ plots again for experimental parameters 
$h_0 = 5\pi$ and ${\cal J} = h_0/20$ (in rad/s). 
The peaks represent the freezing points.
For numerical simulation of 3-spin Ising chain, we observe three freezing points in this
range, viz. at $ \omega =$ 3.59, 5.61, and 12.88 rad/s.
These points are very close to the one predicted by analytical formula (Eq.~\ref{fc}),
which gives freezing values to be $ \omega =$ 3.63, 5.69, and 13.06 rad/s in our case.
Comparison of the experimental data after inverse-decay correction
with the exact numerical simulation for the $Q$-lines
exhibits striking agreement. 
The raw experimental data exhibits an over-all downward shift
of the experimental line due to the $T_{d}$ decay, though the basic features (particularly, the peak positions) 
remain same. We also compare the finite size result with the infinite size analytical formula (\ref{Q_Ana})
in Fig.~\ref{qplot}.
\noindent Interestingly, not only the freezing peaks, 
but also the trend of most of the $Q$ vs $\omega$ profile 
for our 3-spin system driven over 30 cycles matches fairly well with that of 
an infinite chain driven over infinite time! 
In Fig.~\ref{spec}, we show the experimentally obtained spectra 
for $ m^x(0)=1 $ at time instants $0,3\tau,6\tau,\cdots,30\tau$,
for two driving frequencies (i) $\omega=24.54$ rad/s 
not satisfying the freezing condition (left column)
and (ii) $\omega = 5.61$ rad/s satisfying the freezing
condition (right column).
The spectra at time instant $ n=0 $ correspond to the state completely polarized in $+x$ direction.	
The phase-oscillations of the 
the spectral lines in the non-freezing case (left column) can
be clearly noticed, while the freezing spectra (right column)
remain in-phase.

\subsection{Role of Quantum Interference: Amplitude Vs Probability}
\begin{center}
\begin{figure}[h]
\centering
\includegraphics[width=0.9\linewidth]{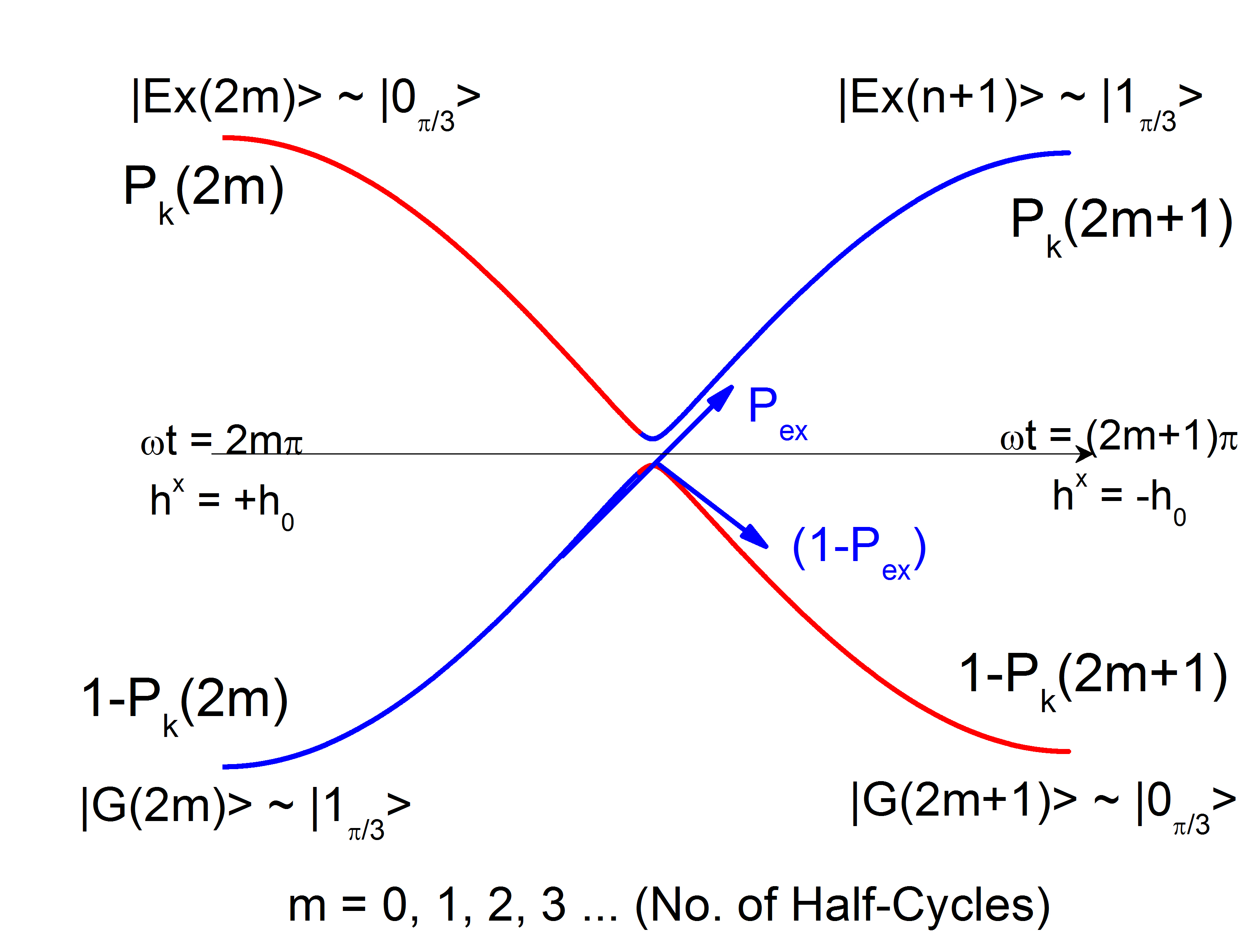}
\caption{(Color online) Plot of energies of the instantaneous ground state (denoted by $|G\rangle$) and excited state ($|Ex\rangle$)
of the $2\times 2$ Hamiltonian (Eq.~\ref{H2}) as a function of $\omega t$. Our first sweep $(m=0)$
starts from the left end with the initial state $|\psi(0)\rangle_{\pi/3} = \{u_{\pi/3}(0)=0;v_{\pi/3}(0)=1\} \approx |G(0)\rangle$.
$P_{ex}$ denotes the probability that starting from this initial state the system ends up with the same state
(i.e. $\{u_{\pi/3}=0;v_{\pi/3}=1\}$) at the end of half a cycle, instead of following the drive adiabatically
and ending up with the ground state ($\approx |0_{\pi/3}\rangle$) of the final Hamiltonian.
}
\label{LZ-Cartoon}
\end{figure}
\end{center}
It is interesting to consider the crucial role of quantum interference in this phenomenon.
The non-monotonic nature of the freezing phenomena, including the appearance of maximal freezing peaks
are consequence of quantum interference, as indicated in \cite{AD-DQH}.
Here we explicitly demonstrate the role of quantum interference behind the key features of the
freezing behavior.
To this end we analyze the problem using another approach based on repeated calculation of Landau-Zener like excitation,
which has been employed successfully to estimate certain aspects
of a repeated quench dynamics in a similar model in presence of decoherence after each sweep \cite{Amit-deco}.
Role of quantum interference in repeated quenches has been studied earlier \cite{victor}.
We demonstrate how such an approach (based on counting of probabilities rather than amplitudes)
fails to explain even the qualitative features of the freezing phenomenon.
Suppose we start from
the initial state $\{u_{\pi/3}(0) \approx 0, v_{\pi/3}(0)\approx 1\}$, and make $2m$ number of half-cycles
and find that we are in a state $\{u_{\pi/3}(2m), v_{\pi/3}(2m)\}$, where $P_{k}(2m) = |v_{\pi/3}(2m)|^2$
(see Fig.~\ref{LZ-Cartoon}).
Now we want to estimate the probability $|v_{\pi/3}(2m+1)|^3$ of being in the state
$|1_{\pi/3}\rangle$ after the subsequent (i.e. $2m+1$-th) half-sweep is made. We can come up with
an estimate based on calculating transition probabilities if we are supplied with the excitation probability $P_{ex}$,
which is the probability that we start with the states $|1_{\pi/3}\rangle$ before starting a half-cycle and
end up with the same state $|1_{\pi/3}\rangle$ after completing the half-cycle. The name excitation probability
is motivated by the fact that since $h_{0} \gg {\cal J}$, for $h^{x} = +h_{0}$ the ground state
(to a good approximation) corresponds to $|1_{\pi/3}\rangle$ (left end of Fig.~\ref{LZ-Cartoon}),
while for $h^{x} = -h_{0}$ the ground state approximately
corresponds to  $|0_{\pi/3}\rangle$ (right end of the Fig.).
Due to the symmetry of the
problem, this is also the probability that we end up in the ground state starting from the excited state.
Since  $h_{0} \gg {\cal J}\cos{k}$, we can approximately use the same $P_{ex}$ for the reverse sweep
(using a different one for the reverse sweep doesn't change the conclusion in any qualitative way).
Now taking the excitation probabilities into account we find the following recursion relation
\be
|v_{\pi/3}(2m+1)|^2 = |v_{\pi/3}(2m)|^2(2P_{ex} - 1) + 1-P_{ex},
\label{recursion}
\ee
\noindent
where $m=0$ denotes the initial state (starting from the left).
Solving above relation one gets
\be
|v_{\pi/3}(2m+1)|^{2} = \frac{1}{2} + (2P_{ex} - 1)^{2m}[|v_{\pi/3}(0)|^2 -\frac{1}{2}]
\label{Rec-Sol}
\ee
\noindent
Thus $|v_{\pi/3}(2m+1)|^{2} \to \frac{1}{2}$ as $m \to \infty.$ 
This implies $Q \to 0$ {\it regardless of} $\omega$ as $m \to \infty$ as $L\to\infty.$
In our case ($L = 3$) this would give (using Eq.~\ref{mx-3}),
\be
\label{Q-3-strobo}
\lim_{N_{s}\to\infty}Q_{Prob}(N_{s}) = \lim_{N_{s}\to\infty}\frac{1}{N_{s}}\sum^{N_{s}}_{n=0}m^x(t=n\tau) = \frac{1}{3},
\ee
\noindent
\begin{center}
\begin{figure}[ht]
\centering
\includegraphics[width=0.9\linewidth]{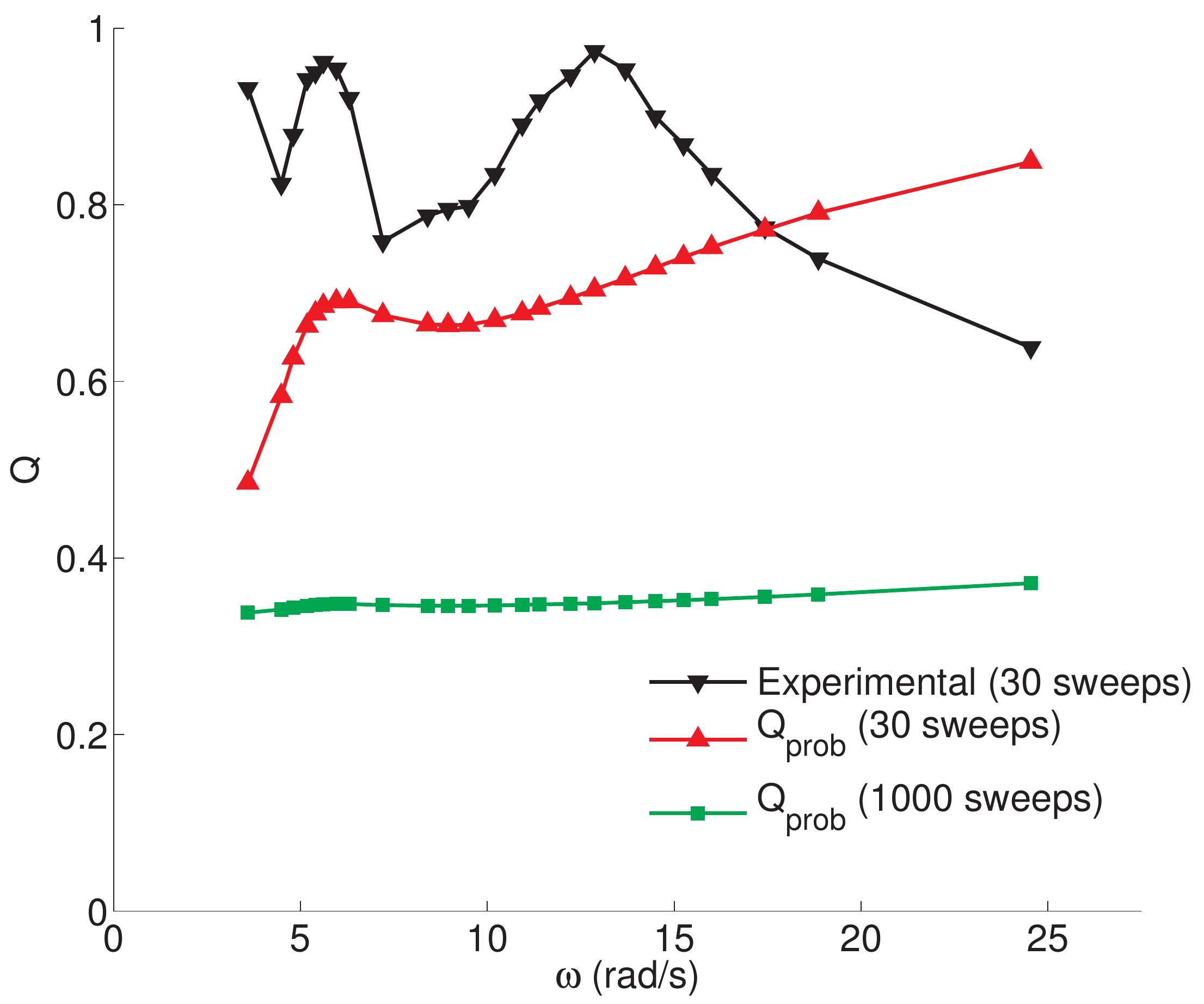}
\caption{(Color online) Qualitative feature of the theory that uses only transition probabilities
between the states $|u_{\pi/3}\rangle,|v_{\pi/3}\rangle$ after each half-cycle,
neglecting the phase between them. This illustrates the importance of quantum interference --
the role those phases play in shaping the qualitative characteristics of the freezing phenomenon.
The full quantum treatment provides excellent agreement with the experimental data (Fig~\ref{qplot}).
The results correspond to experimental parameter values: $h_{0} = 5\pi$, ${\cal {J}} = h_{0}/20$ (in rad/s).
}
\label{P-vs-A}
\end{figure}
\end{center}
where $Q_{Prob}$ denotes $Q$ calculated using transition probabilities (instead of amplitudes) as described above. 
This shows that the variation of $Q$ with $\omega$ and $h_{0}$
is itself a result of repeated quantum interference between the phases gathered after the half cycles.

We see a qualitative difference between the experimental results
(which agrees with the correct quantum treatment, as shown in Fig.~\ref{qplot}) 
and the above probability based calculation. 
In Fig.~\ref{P-vs-A} we show an explicit comparison of 
the experimental result with the numerical result calculated using Eqs.~(\ref{Rec-Sol} and~\ref{mx-3})
for the experimental parameter values. $Q_{Prob}$ is also calculated stroboscopically 
over $30$ cycles as has been done in the experiment. $P_{ex}$ is calculated exactly by 
evolving the system numerically for half-a-cycle for each value of $\omega$.
$m^x(n)$ is measured after each complete cycle and averaged over cycles to calculate $Q_{Prob}$ (as is done in the experiment).
The result shows a monotonic behavior for $Q_{Prob}$ for 30 cycles (orange line-triangles) as a function of $\omega$ for large 
$\omega$, only with a reminiscence of the non-monotonic behavior close to 
the first peak. However, for $Q_{Prob}$ averaged over a larger number 
of sweeps this profile tends to get flatten out completely, as demonstrated by the 
plot of $Q_{Prob}$ for 1000 sweeps (green line-squares). 
In that case $Q \approx 3$ independent of $\omega$ as predicted by Eq.(~\ref{Q-3-strobo}).
The probability based calculations that neglects the phase between $|u_{\pi/3}\rangle$
and $|v_{\pi/3}\rangle$, thus produces a qualitatively wrong result, even within our experimental window
of time and small system-size. This illustrates the significance of quantum interference in the
freezing phenomenon.  

\subsection{Conclusion and Outlook} 
In our NMR experiments we have demonstrated that repeated quantum interference 
can strongly freeze the magnetization dynamics in
of a periodically driven Ising chain for certain particular values
of the driving amplitude and the frequency, and confirmed the phenomenon of
dynamical many-body freezing with excellent quantitative accuracy for both initial states. 
This is a major step forward, where a surprising non-equilibrium manifestation of 
quantum interference has been accurately demonstrated experimentally. 
With further experimental accuracies, complete freezing of certain quantities against unwanted evolution 
in a system of interacting qubits (Ising spins) might be possible by imposing  
strong Ising interactions between the qubits and a suitable periodic drive.   
Such a coherent control mechanism might inspire useful technologies as it has been done in the past.
For example, the average Hamiltonian theory developed by Waugh and 
co-workers, founded high-resolution solid state Nuclear Magnetic Resonance (NMR)  \cite{haeberlen}.
It involves coherent driving of spin systems via a time-dependent
Hamiltonian, leading to an effective evolution under desired
interactions.  A constant drive with a strong transverse field, called
spin-lock, is used in many of the NMR experiments to create an effective
transverse Hamiltonian in the rotating frame \cite{cavanagh1995protein}.


The authors are grateful to B. K. Chakrabarti, A. Kumar, A. Lazarides, R. Moessner and  
S. S. Roy for useful discussions. This work was partly supported by the DST 
Project SR/S2/LOP-0017/2009.

\bibliographystyle{apsrev4-1}
\bibliography{bib_freezing}

\begin{thebibliography}{55}%
\makeatletter
\providecommand \@ifxundefined [1]{%
 \@ifx{#1\undefined}
}%
\providecommand \@ifnum [1]{%
 \ifnum #1\expandafter \@firstoftwo
 \else \expandafter \@secondoftwo
 \fi
}%
\providecommand \@ifx [1]{%
 \ifx #1\expandafter \@firstoftwo
 \else \expandafter \@secondoftwo
 \fi
}%
\providecommand \natexlab [1]{#1}%
\providecommand \enquote  [1]{``#1''}%
\providecommand \bibnamefont  [1]{#1}%
\providecommand \bibfnamefont [1]{#1}%
\providecommand \citenamefont [1]{#1}%
\providecommand \href@noop [0]{\@secondoftwo}%
\providecommand \href [0]{\begingroup \@sanitize@url \@href}%
\providecommand \@href[1]{\@@startlink{#1}\@@href}%
\providecommand \@@href[1]{\endgroup#1\@@endlink}%
\providecommand \@sanitize@url [0]{\catcode `\\12\catcode `\$12\catcode
  `\&12\catcode `\#12\catcode `\^12\catcode `\_12\catcode `\%12\relax}%
\providecommand \@@startlink[1]{}%
\providecommand \@@endlink[0]{}%
\providecommand \url  [0]{\begingroup\@sanitize@url \@url }%
\providecommand \@url [1]{\endgroup\@href {#1}{\urlprefix }}%
\providecommand \urlprefix  [0]{URL }%
\providecommand \Eprint [0]{\href }%
\providecommand \doibase [0]{http://dx.doi.org/}%
\providecommand \selectlanguage [0]{\@gobble}%
\providecommand \bibinfo  [0]{\@secondoftwo}%
\providecommand \bibfield  [0]{\@secondoftwo}%
\providecommand \translation [1]{[#1]}%
\providecommand \BibitemOpen [0]{}%
\providecommand \bibitemStop [0]{}%
\providecommand \bibitemNoStop [0]{.\EOS\space}%
\providecommand \EOS [0]{\spacefactor3000\relax}%
\providecommand \BibitemShut  [1]{\csname bibitem#1\endcsname}%
\let\auto@bib@innerbib\@empty
\bibitem [{\citenamefont {Tseng~\textit{et. al.}}(2000)}]{NMR-1}%
  \BibitemOpen
  \bibfield  {author} {\bibinfo {author} {\bibfnamefont {C.~H.}\ \bibnamefont
  {Tseng~\textit{et. al.}}},\ }\href {\doibase 10.1103/PhysRevA.62.032309}
  {\bibfield  {journal} {\bibinfo  {journal} {Phys. Rev. A}\ }\textbf {\bibinfo
  {volume} {62}},\ \bibinfo {pages} {032309} (\bibinfo {year}
  {2000})}\BibitemShut {NoStop}%
\bibitem [{\citenamefont {Wu}\ \emph {et~al.}(2002)\citenamefont {Wu},
  \citenamefont {Byrd},\ and\ \citenamefont {Lidar}}]{NMR-2}%
  \BibitemOpen
  \bibfield  {author} {\bibinfo {author} {\bibfnamefont {L.-A.}\ \bibnamefont
  {Wu}}, \bibinfo {author} {\bibfnamefont {M.~S.}\ \bibnamefont {Byrd}}, \ and\
  \bibinfo {author} {\bibfnamefont {D.~A.}\ \bibnamefont {Lidar}},\ }\href
  {\doibase 10.1103/PhysRevLett.89.057904} {\bibfield  {journal} {\bibinfo
  {journal} {Phys. Rev. Lett.}\ }\textbf {\bibinfo {volume} {89}},\ \bibinfo
  {pages} {057904} (\bibinfo {year} {2002})}\BibitemShut {NoStop}%
\bibitem [{\citenamefont {Weinstein}\ \emph {et~al.}(2002)\citenamefont
  {Weinstein}, \citenamefont {Lloyd}, \citenamefont {Emerson},\ and\
  \citenamefont {Cory}}]{NMR-3}%
  \BibitemOpen
  \bibfield  {author} {\bibinfo {author} {\bibfnamefont {Y.~S.}\ \bibnamefont
  {Weinstein}}, \bibinfo {author} {\bibfnamefont {S.}~\bibnamefont {Lloyd}},
  \bibinfo {author} {\bibfnamefont {J.}~\bibnamefont {Emerson}}, \ and\
  \bibinfo {author} {\bibfnamefont {D.~G.}\ \bibnamefont {Cory}},\ }\href
  {\doibase 10.1103/PhysRevLett.89.157902} {\bibfield  {journal} {\bibinfo
  {journal} {Phys. Rev. Lett.}\ }\textbf {\bibinfo {volume} {89}},\ \bibinfo
  {pages} {157902} (\bibinfo {year} {2002})}\BibitemShut {NoStop}%
\bibitem [{\citenamefont {Negrevergne~\textit{et. al.}}(2005)}]{NMR-4}%
  \BibitemOpen
  \bibfield  {author} {\bibinfo {author} {\bibfnamefont {C.}~\bibnamefont
  {Negrevergne~\textit{et. al.}}},\ }\href {\doibase
  10.1103/PhysRevA.71.032344} {\bibfield  {journal} {\bibinfo  {journal} {Phys.
  Rev. A}\ }\textbf {\bibinfo {volume} {71}},\ \bibinfo {pages} {032344}
  (\bibinfo {year} {2005})}\BibitemShut {NoStop}%
\bibitem [{\citenamefont {Brown}\ \emph {et~al.}(2006)\citenamefont {Brown},
  \citenamefont {Clark},\ and\ \citenamefont {Chuang}}]{NMR-5}%
  \BibitemOpen
  \bibfield  {author} {\bibinfo {author} {\bibfnamefont {K.~R.}\ \bibnamefont
  {Brown}}, \bibinfo {author} {\bibfnamefont {R.~J.}\ \bibnamefont {Clark}}, \
  and\ \bibinfo {author} {\bibfnamefont {I.~L.}\ \bibnamefont {Chuang}},\
  }\href {\doibase 10.1103/PhysRevLett.97.050504} {\bibfield  {journal}
  {\bibinfo  {journal} {Phys. Rev. Lett.}\ }\textbf {\bibinfo {volume} {97}},\
  \bibinfo {pages} {050504} (\bibinfo {year} {2006})}\BibitemShut {NoStop}%
\bibitem [{\citenamefont {Yang}\ \emph {et~al.}(2006)\citenamefont {Yang},
  \citenamefont {Wang}, \citenamefont {Xu},\ and\ \citenamefont {Du}}]{NMR-6}%
  \BibitemOpen
  \bibfield  {author} {\bibinfo {author} {\bibfnamefont {X.}~\bibnamefont
  {Yang}}, \bibinfo {author} {\bibfnamefont {A.~M.}\ \bibnamefont {Wang}},
  \bibinfo {author} {\bibfnamefont {F.}~\bibnamefont {Xu}}, \ and\ \bibinfo
  {author} {\bibfnamefont {J.}~\bibnamefont {Du}},\ }\href {\doibase
  10.1016/j.cplett.2006.02.023} {\bibfield  {journal} {\bibinfo  {journal}
  {Chemical Physics Letters}\ }\textbf {\bibinfo {volume} {422}},\ \bibinfo
  {pages} {20 } (\bibinfo {year} {2006})}\BibitemShut {NoStop}%
\bibitem [{\citenamefont {Roumpos}\ \emph {et~al.}(2007)\citenamefont
  {Roumpos}, \citenamefont {Master},\ and\ \citenamefont {Yamamoto}}]{NMR-7}%
  \BibitemOpen
  \bibfield  {author} {\bibinfo {author} {\bibfnamefont {G.}~\bibnamefont
  {Roumpos}}, \bibinfo {author} {\bibfnamefont {C.~P.}\ \bibnamefont {Master}},
  \ and\ \bibinfo {author} {\bibfnamefont {Y.}~\bibnamefont {Yamamoto}},\
  }\href {\doibase 10.1103/PhysRevB.75.094415} {\bibfield  {journal} {\bibinfo
  {journal} {Phys. Rev. B}\ }\textbf {\bibinfo {volume} {75}},\ \bibinfo
  {pages} {094415} (\bibinfo {year} {2007})}\BibitemShut {NoStop}%
\bibitem [{\citenamefont {Bloch}\ \emph {et~al.}(2012)\citenamefont {Bloch},
  \citenamefont {Dalibard},\ and\ \citenamefont {Nascimb{\`e}ne}}]{Bloch-1}%
  \BibitemOpen
  \bibfield  {author} {\bibinfo {author} {\bibfnamefont {I.}~\bibnamefont
  {Bloch}}, \bibinfo {author} {\bibfnamefont {J.}~\bibnamefont {Dalibard}}, \
  and\ \bibinfo {author} {\bibfnamefont {S.}~\bibnamefont {Nascimb{\`e}ne}},\
  }\href {http://dx.doi.org/10.1038/nphys2259} {\bibfield  {journal} {\bibinfo
  {journal} {Nature Physics}\ }\textbf {\bibinfo {volume} {8}},\ \bibinfo
  {pages} {267} (\bibinfo {year} {2012})}\BibitemShut {NoStop}%
\bibitem [{\citenamefont {Bloch}\ \emph {et~al.}(2008)\citenamefont {Bloch},
  \citenamefont {Dalibard},\ and\ \citenamefont {Zwerger}}]{Bloch-2}%
  \BibitemOpen
  \bibfield  {author} {\bibinfo {author} {\bibfnamefont {I.}~\bibnamefont
  {Bloch}}, \bibinfo {author} {\bibfnamefont {J.}~\bibnamefont {Dalibard}}, \
  and\ \bibinfo {author} {\bibfnamefont {W.}~\bibnamefont {Zwerger}},\ }\href
  {\doibase 10.1103/RevModPhys.80.885} {\bibfield  {journal} {\bibinfo
  {journal} {Rev. Mod. Phys.}\ }\textbf {\bibinfo {volume} {80}},\ \bibinfo
  {pages} {885} (\bibinfo {year} {2008})}\BibitemShut {NoStop}%
\bibitem [{\citenamefont {Lanyon~\textit{et. al. }}(2011)}]{Blatt}%
  \BibitemOpen
  \bibfield  {author} {\bibinfo {author} {\bibfnamefont {B.~P.}\ \bibnamefont
  {Lanyon~\textit{et. al. }}},\ }\href {\doibase 10.1126/science.1208001}
  {\bibfield  {journal} {\bibinfo  {journal} {Science}\ }\textbf {\bibinfo
  {volume} {334}},\ \bibinfo {pages} {57} (\bibinfo {year} {2011})}\BibitemShut
  {NoStop}%
\bibitem [{\citenamefont {Eckardt}\ \emph {et~al.}(2005)\citenamefont
  {Eckardt}, \citenamefont {Weiss},\ and\ \citenamefont {Holthaus}}]{Andre-1}%
  \BibitemOpen
  \bibfield  {author} {\bibinfo {author} {\bibfnamefont {A.}~\bibnamefont
  {Eckardt}}, \bibinfo {author} {\bibfnamefont {C.}~\bibnamefont {Weiss}}, \
  and\ \bibinfo {author} {\bibfnamefont {M.}~\bibnamefont {Holthaus}},\ }\href
  {\doibase 10.1103/PhysRevLett.95.260404} {\bibfield  {journal} {\bibinfo
  {journal} {Phys. Rev. Lett.}\ }\textbf {\bibinfo {volume} {95}},\ \bibinfo
  {pages} {260404} (\bibinfo {year} {2005})}\BibitemShut {NoStop}%
\bibitem [{\citenamefont {Eckardt}\ and\ \citenamefont
  {Holthaus}(2008)}]{Andre-2}%
  \BibitemOpen
  \bibfield  {author} {\bibinfo {author} {\bibfnamefont {A.}~\bibnamefont
  {Eckardt}}\ and\ \bibinfo {author} {\bibfnamefont {M.}~\bibnamefont
  {Holthaus}},\ }\href {\doibase 10.1103/PhysRevLett.101.245302} {\bibfield
  {journal} {\bibinfo  {journal} {Phys. Rev. Lett.}\ }\textbf {\bibinfo
  {volume} {101}},\ \bibinfo {pages} {245302} (\bibinfo {year}
  {2008})}\BibitemShut {NoStop}%
\bibitem [{\citenamefont {Lazarides}\ \emph
  {et~al.}(2014{\natexlab{a}})\citenamefont {Lazarides}, \citenamefont {Das},\
  and\ \citenamefont {Moessner}}]{AAR-PRL}%
  \BibitemOpen
  \bibfield  {author} {\bibinfo {author} {\bibfnamefont {A.}~\bibnamefont
  {Lazarides}}, \bibinfo {author} {\bibfnamefont {A.}~\bibnamefont {Das}}, \
  and\ \bibinfo {author} {\bibfnamefont {R.}~\bibnamefont {Moessner}},\ }\href
  {\doibase 10.1103/PhysRevLett.112.150401} {\bibfield  {journal} {\bibinfo
  {journal} {Phys. Rev. Letts.}\ }\textbf {\bibinfo {volume} {112}},\ \bibinfo
  {pages} {150401} (\bibinfo {year} {2014}{\natexlab{a}})}\BibitemShut
  {NoStop}%
\bibitem [{\citenamefont {Lazarides}\ \emph
  {et~al.}(2014{\natexlab{b}})\citenamefont {Lazarides}, \citenamefont {Das},\
  and\ \citenamefont {Moessner}}]{AAR-Generic}%
  \BibitemOpen
  \bibfield  {author} {\bibinfo {author} {\bibfnamefont {A.}~\bibnamefont
  {Lazarides}}, \bibinfo {author} {\bibfnamefont {A.}~\bibnamefont {Das}}, \
  and\ \bibinfo {author} {\bibfnamefont {R.}~\bibnamefont {Moessner}},\
  }\href@noop {} {\bibfield  {journal} {\bibinfo  {journal} {arXiv:1403.2946}\
  }\textbf {\bibinfo {volume} {112}} (\bibinfo {year}
  {2014}{\natexlab{b}})}\BibitemShut {NoStop}%
\bibitem [{\citenamefont {Arimondo~\textit{et. al.}}(2012)}]{Arimondo-Rev}%
  \BibitemOpen
  \bibfield  {author} {\bibinfo {author} {\bibfnamefont {E.}~\bibnamefont
  {Arimondo~\textit{et. al.}}},\ }in\ \href {\doibase
  10.1016/B978-0-12-396482-3.00010-7} {\emph {\bibinfo {booktitle} {Advances in
  Atomic, Molecular, and Optical Physics}}},\ Vol.~\bibinfo {volume} {61}\
  (\bibinfo  {publisher} {Academic Press},\ \bibinfo {year} {2012})\BibitemShut
  {NoStop}%
\bibitem [{\citenamefont {Struck~\textit{et. al.}}(2012)}]{Gauge-1a}%
  \BibitemOpen
  \bibfield  {author} {\bibinfo {author} {\bibfnamefont {J.}~\bibnamefont
  {Struck~\textit{et. al.}}},\ }\href {\doibase 10.1103/PhysRevLett.108.225304}
  {\bibfield  {journal} {\bibinfo  {journal} {Phys. Rev. Lett.}\ }\textbf
  {\bibinfo {volume} {108}},\ \bibinfo {pages} {225304} (\bibinfo {year}
  {2012})}\BibitemShut {NoStop}%
\bibitem [{\citenamefont {Hauke~\textit{et.al.}}(2012)}]{Gauge-1b}%
  \BibitemOpen
  \bibfield  {author} {\bibinfo {author} {\bibfnamefont {P.}~\bibnamefont
  {Hauke~\textit{et.al.}}},\ }\href {\doibase 10.1103/PhysRevLett.109.145301}
  {\bibfield  {journal} {\bibinfo  {journal} {Phys. Rev. Lett.}\ }\textbf
  {\bibinfo {volume} {109}},\ \bibinfo {pages} {145301} (\bibinfo {year}
  {2012})}\BibitemShut {NoStop}%
\bibitem [{\citenamefont {Lindner}\ \emph {et~al.}(2011)\citenamefont
  {Lindner}, \citenamefont {Refael},\ and\ \citenamefont {Galitski}}]{Gauge-2}%
  \BibitemOpen
  \bibfield  {author} {\bibinfo {author} {\bibfnamefont {N.~H.}\ \bibnamefont
  {Lindner}}, \bibinfo {author} {\bibfnamefont {G.}~\bibnamefont {Refael}}, \
  and\ \bibinfo {author} {\bibfnamefont {V.}~\bibnamefont {Galitski}},\ }\href
  {http://dx.doi.org/10.1038/nphys1926} {\bibfield  {journal} {\bibinfo
  {journal} {Nature Physics}\ }\textbf {\bibinfo {volume} {7}},\ \bibinfo
  {pages} {490} (\bibinfo {year} {2011})}\BibitemShut {NoStop}%
\bibitem [{\citenamefont {Prosen}\ and\ \citenamefont
  {Ilievski}(2011)}]{Prosen-1}%
  \BibitemOpen
  \bibfield  {author} {\bibinfo {author} {\bibfnamefont {T.}~\bibnamefont
  {Prosen}}\ and\ \bibinfo {author} {\bibfnamefont {E.}~\bibnamefont
  {Ilievski}},\ }\href {\doibase 10.1103/PhysRevLett.107.060403} {\bibfield
  {journal} {\bibinfo  {journal} {Phys. Rev. Lett.}\ }\textbf {\bibinfo
  {volume} {107}},\ \bibinfo {pages} {060403} (\bibinfo {year}
  {2011})}\BibitemShut {NoStop}%
\bibitem [{\citenamefont {{Mondal, S.}}\ \emph {et~al.}(2012)\citenamefont
  {{Mondal, S.}}, \citenamefont {{Pekker, D.}},\ and\ \citenamefont {{Sengupta,
  K.}}}]{Kris-Periodic}%
  \BibitemOpen
  \bibfield  {author} {\bibinfo {author} {\bibnamefont {{Mondal, S.}}},
  \bibinfo {author} {\bibnamefont {{Pekker, D.}}}, \ and\ \bibinfo {author}
  {\bibnamefont {{Sengupta, K.}}},\ }\href {\doibase
  10.1209/0295-5075/100/60007} {\bibfield  {journal} {\bibinfo  {journal}
  {EPL}\ }\textbf {\bibinfo {volume} {100}},\ \bibinfo {pages} {60007}
  (\bibinfo {year} {2012})}\BibitemShut {NoStop}%
\bibitem [{\citenamefont {Bastidas}\ \emph {et~al.}(2012)\citenamefont
  {Bastidas}, \citenamefont {Emary}, \citenamefont {Schaller},\ and\
  \citenamefont {Brandes}}]{Bastidas}%
  \BibitemOpen
  \bibfield  {author} {\bibinfo {author} {\bibfnamefont {V.~M.}\ \bibnamefont
  {Bastidas}}, \bibinfo {author} {\bibfnamefont {C.}~\bibnamefont {Emary}},
  \bibinfo {author} {\bibfnamefont {G.}~\bibnamefont {Schaller}}, \ and\
  \bibinfo {author} {\bibfnamefont {T.}~\bibnamefont {Brandes}},\ }\href
  {\doibase 10.1103/PhysRevA.86.063627} {\bibfield  {journal} {\bibinfo
  {journal} {Phys. Rev. A}\ }\textbf {\bibinfo {volume} {86}},\ \bibinfo
  {pages} {063627} (\bibinfo {year} {2012})}\BibitemShut {NoStop}%
\bibitem [{\citenamefont {Mukherjee}\ and\ \citenamefont
  {Dutta}(2009)}]{victor}%
  \BibitemOpen
  \bibfield  {author} {\bibinfo {author} {\bibfnamefont {V.}~\bibnamefont
  {Mukherjee}}\ and\ \bibinfo {author} {\bibfnamefont {A.}~\bibnamefont
  {Dutta}},\ }\href {http://stacks.iop.org/1742-5468/2009/i=05/a=P05005}
  {\bibfield  {journal} {\bibinfo  {journal} {Journal of Statistical Mechanics:
  Theory and Experiment}\ }\textbf {\bibinfo {volume} {2009}},\ \bibinfo
  {pages} {P05005} (\bibinfo {year} {2009})}\BibitemShut {NoStop}%
\bibitem [{\citenamefont {Nag}\ \emph {et~al.}(2014)\citenamefont {Nag},
  \citenamefont {Roy}, \citenamefont {Dutta},\ and\ \citenamefont
  {Sen}}]{Sthitadhi}%
  \BibitemOpen
  \bibfield  {author} {\bibinfo {author} {\bibfnamefont {T.}~\bibnamefont
  {Nag}}, \bibinfo {author} {\bibfnamefont {S.}~\bibnamefont {Roy}}, \bibinfo
  {author} {\bibfnamefont {A.}~\bibnamefont {Dutta}}, \ and\ \bibinfo {author}
  {\bibfnamefont {D.}~\bibnamefont {Sen}},\ }\href@noop {} {\bibfield
  {journal} {\bibinfo  {journal} {arXiv:1312.6467 (accepted in Phys. Rev. B.)}\
  } (\bibinfo {year} {2014})}\BibitemShut {NoStop}%
\bibitem [{\citenamefont {Lignier~\textit{et. al.}}(2007)}]{Arimondo-1}%
  \BibitemOpen
  \bibfield  {author} {\bibinfo {author} {\bibfnamefont {H.}~\bibnamefont
  {Lignier~\textit{et. al.}}},\ }\href {\doibase 10.1103/PhysRevLett.99.220403}
  {\bibfield  {journal} {\bibinfo  {journal} {Phys. Rev. Lett.}\ }\textbf
  {\bibinfo {volume} {99}},\ \bibinfo {pages} {220403} (\bibinfo {year}
  {2007})}\BibitemShut {NoStop}%
\bibitem [{\citenamefont {Zenesini~\textit{et. al.}}(2009)}]{Arimondo-2}%
  \BibitemOpen
  \bibfield  {author} {\bibinfo {author} {\bibfnamefont {A.}~\bibnamefont
  {Zenesini~\textit{et. al.}}},\ }\href {\doibase
  10.1103/PhysRevLett.102.100403} {\bibfield  {journal} {\bibinfo  {journal}
  {Phys. Rev. Lett.}\ }\textbf {\bibinfo {volume} {102}},\ \bibinfo {pages}
  {100403} (\bibinfo {year} {2009})}\BibitemShut {NoStop}%
\bibitem [{\citenamefont {Chen}\ \emph {et~al.}(2011)\citenamefont {Chen},
  \citenamefont {Nascimb\`ene}, \citenamefont {Aidelsburger}, \citenamefont
  {Atala}, \citenamefont {Trotzky},\ and\ \citenamefont
  {Bloch}}]{Bloch-Periodic}%
  \BibitemOpen
  \bibfield  {author} {\bibinfo {author} {\bibfnamefont {Y.-A.}\ \bibnamefont
  {Chen}}, \bibinfo {author} {\bibfnamefont {S.}~\bibnamefont {Nascimb\`ene}},
  \bibinfo {author} {\bibfnamefont {M.}~\bibnamefont {Aidelsburger}}, \bibinfo
  {author} {\bibfnamefont {M.}~\bibnamefont {Atala}}, \bibinfo {author}
  {\bibfnamefont {S.}~\bibnamefont {Trotzky}}, \ and\ \bibinfo {author}
  {\bibfnamefont {I.}~\bibnamefont {Bloch}},\ }\href {\doibase
  10.1103/PhysRevLett.107.210405} {\bibfield  {journal} {\bibinfo  {journal}
  {Phys. Rev. Lett.}\ }\textbf {\bibinfo {volume} {107}},\ \bibinfo {pages}
  {210405} (\bibinfo {year} {2011})}\BibitemShut {NoStop}%
\bibitem [{\citenamefont {Struck~\textit{et. al.}}(2011)}]{Shengstock}%
  \BibitemOpen
  \bibfield  {author} {\bibinfo {author} {\bibfnamefont {J.}~\bibnamefont
  {Struck~\textit{et. al.}}},\ }\href {\doibase 10.1126/science.1207239}
  {\bibfield  {journal} {\bibinfo  {journal} {Science}\ }\textbf {\bibinfo
  {volume} {333}},\ \bibinfo {pages} {996} (\bibinfo {year}
  {2011})}\BibitemShut {NoStop}%
\bibitem [{\citenamefont {Haeberlen}(1976)}]{haeberlen}%
  \BibitemOpen
  \bibfield  {author} {\bibinfo {author} {\bibfnamefont {U.}~\bibnamefont
  {Haeberlen}},\ }\href@noop {} {\emph {\bibinfo {title} {High resolution NMR
  in solids: selective averaging}}},\ Vol.~\bibinfo {volume} {1}\ (\bibinfo
  {publisher} {Academic Press New York},\ \bibinfo {year} {1976})\BibitemShut
  {NoStop}%
\bibitem [{\citenamefont {Cavanagh}\ \emph {et~al.}(1995)\citenamefont
  {Cavanagh}, \citenamefont {Fairbrother}, \citenamefont {Palmer~III},\ and\
  \citenamefont {Skelton}}]{cavanagh1995protein}%
  \BibitemOpen
  \bibfield  {author} {\bibinfo {author} {\bibfnamefont {J.}~\bibnamefont
  {Cavanagh}}, \bibinfo {author} {\bibfnamefont {W.~J.}\ \bibnamefont
  {Fairbrother}}, \bibinfo {author} {\bibfnamefont {A.~G.}\ \bibnamefont
  {Palmer~III}}, \ and\ \bibinfo {author} {\bibfnamefont {N.~J.}\ \bibnamefont
  {Skelton}},\ }\href@noop {} {\emph {\bibinfo {title} {Protein NMR
  spectroscopy: principles and practice}}}\ (\bibinfo  {publisher} {Academic
  Press},\ \bibinfo {year} {1995})\BibitemShut {NoStop}%
\bibitem [{\citenamefont {Das}(2010)}]{AD-DQH}%
  \BibitemOpen
  \bibfield  {author} {\bibinfo {author} {\bibfnamefont {A.}~\bibnamefont
  {Das}},\ }\href {\doibase 10.1103/PhysRevB.82.172402} {\bibfield  {journal}
  {\bibinfo  {journal} {Phys. Rev. B}\ }\textbf {\bibinfo {volume} {82}},\
  \bibinfo {pages} {172402} (\bibinfo {year} {2010})}\BibitemShut {NoStop}%
\bibitem [{\citenamefont {Kibble}(1976)}]{Kibble}%
  \BibitemOpen
  \bibfield  {author} {\bibinfo {author} {\bibfnamefont {T.~W.~B.}\
  \bibnamefont {Kibble}},\ }\href {http://stacks.iop.org/0305-4470/9/i=8/a=029}
  {\bibfield  {journal} {\bibinfo  {journal} {Journal of Physics A:
  Mathematical and General}\ }\textbf {\bibinfo {volume} {9}},\ \bibinfo
  {pages} {1387} (\bibinfo {year} {1976})}\BibitemShut {NoStop}%
\bibitem [{\citenamefont {Zurek}(1985)}]{Zurek}%
  \BibitemOpen
  \bibfield  {author} {\bibinfo {author} {\bibfnamefont {W.}~\bibnamefont
  {Zurek}},\ }\href {http://dx.doi.org/10.1038/317505a0} {\bibfield  {journal}
  {\bibinfo  {journal} {Nature}\ }\textbf {\bibinfo {volume} {317}},\ \bibinfo
  {pages} {505} (\bibinfo {year} {1985})}\BibitemShut {NoStop}%
\bibitem [{\citenamefont {Damski}(2005)}]{Bogdan-LZ}%
  \BibitemOpen
  \bibfield  {author} {\bibinfo {author} {\bibfnamefont {B.}~\bibnamefont
  {Damski}},\ }\href {\doibase 10.1103/PhysRevLett.95.035701} {\bibfield
  {journal} {\bibinfo  {journal} {Phys. Rev. Lett.}\ }\textbf {\bibinfo
  {volume} {95}},\ \bibinfo {pages} {035701} (\bibinfo {year}
  {2005})}\BibitemShut {NoStop}%
\bibitem [{\citenamefont {Zurek}\ \emph {et~al.}(2005)\citenamefont {Zurek},
  \citenamefont {Dorner},\ and\ \citenamefont {Zoller}}]{Zurek-Dorner}%
  \BibitemOpen
  \bibfield  {author} {\bibinfo {author} {\bibfnamefont {W.~H.}\ \bibnamefont
  {Zurek}}, \bibinfo {author} {\bibfnamefont {U.}~\bibnamefont {Dorner}}, \
  and\ \bibinfo {author} {\bibfnamefont {P.}~\bibnamefont {Zoller}},\ }\href
  {\doibase 10.1103/PhysRevLett.95.105701} {\bibfield  {journal} {\bibinfo
  {journal} {Phys. Rev. Lett.}\ }\textbf {\bibinfo {volume} {95}},\ \bibinfo
  {pages} {105701} (\bibinfo {year} {2005})}\BibitemShut {NoStop}%
\bibitem [{\citenamefont {Dziarmaga}(2005)}]{Dziarmaga}%
  \BibitemOpen
  \bibfield  {author} {\bibinfo {author} {\bibfnamefont {J.}~\bibnamefont
  {Dziarmaga}},\ }\href {\doibase 10.1103/PhysRevLett.95.245701} {\bibfield
  {journal} {\bibinfo  {journal} {Phys. Rev. Lett.}\ }\textbf {\bibinfo
  {volume} {95}},\ \bibinfo {pages} {245701} (\bibinfo {year}
  {2005})}\BibitemShut {NoStop}%
\bibitem [{\citenamefont {Landau}\ and\ \citenamefont
  {Lifshitz}(1981)}]{Landau}%
  \BibitemOpen
  \bibfield  {author} {\bibinfo {author} {\bibfnamefont {L.}~\bibnamefont
  {Landau}}\ and\ \bibinfo {author} {\bibfnamefont {E.}~\bibnamefont
  {Lifshitz}},\ }\href@noop {} {\emph {\bibinfo {title} {Quantum mechanics
  non-relativistic theory}}},\ Vol.~\bibinfo {volume} {3}\ (\bibinfo
  {publisher} {Butterworth-Heinemann},\ \bibinfo {year} {1981})\BibitemShut
  {NoStop}%
\bibitem [{\citenamefont {Zener}(1932)}]{Zener}%
  \BibitemOpen
  \bibfield  {author} {\bibinfo {author} {\bibfnamefont {C.}~\bibnamefont
  {Zener}},\ }\href {http://dx.doi.org/10.1098/rspa.1932.0165} {\bibfield
  {journal} {\bibinfo  {journal} {Proceedings of the Royal Society of London.
  Series A, Containing Papers of a Mathematical and Physical Character}\
  }\textbf {\bibinfo {volume} {137}},\ \bibinfo {pages} {696} (\bibinfo {year}
  {1932})}\BibitemShut {NoStop}%
\bibitem [{\citenamefont {Chakrabarti}\ and\ \citenamefont
  {Acharyya}(1999)}]{BKC-RMP}%
  \BibitemOpen
  \bibfield  {author} {\bibinfo {author} {\bibfnamefont {B.~K.}\ \bibnamefont
  {Chakrabarti}}\ and\ \bibinfo {author} {\bibfnamefont {M.}~\bibnamefont
  {Acharyya}},\ }\href {\doibase 10.1103/RevModPhys.71.847} {\bibfield
  {journal} {\bibinfo  {journal} {Rev. Mod. Phys.}\ }\textbf {\bibinfo {volume}
  {71}},\ \bibinfo {pages} {847} (\bibinfo {year} {1999})}\BibitemShut
  {NoStop}%
\bibitem [{\citenamefont {Messiah}(1962)}]{Messiah}%
  \BibitemOpen
  \bibfield  {author} {\bibinfo {author} {\bibfnamefont {A.}~\bibnamefont
  {Messiah}},\ }\href@noop {} {\emph {\bibinfo {title} {Quantum mechanics, vol.
  II}}}\ (\bibinfo  {publisher} {English Ed., North Holland: Amster},\ \bibinfo
  {year} {1962})\BibitemShut {NoStop}%
\bibitem [{\citenamefont {Damski}\ and\ \citenamefont {Zurek}(2006)}]{DZ-LZ}%
  \BibitemOpen
  \bibfield  {author} {\bibinfo {author} {\bibfnamefont {B.}~\bibnamefont
  {Damski}}\ and\ \bibinfo {author} {\bibfnamefont {W.~H.}\ \bibnamefont
  {Zurek}},\ }\href {\doibase 10.1103/PhysRevA.73.063405} {\bibfield  {journal}
  {\bibinfo  {journal} {Phys. Rev. A}\ }\textbf {\bibinfo {volume} {73}},\
  \bibinfo {pages} {063405} (\bibinfo {year} {2006})}\BibitemShut {NoStop}%
\bibitem [{\citenamefont {Bhattacharyya}\ \emph {et~al.}(2012)\citenamefont
  {Bhattacharyya}, \citenamefont {Das},\ and\ \citenamefont
  {Dasgupta}}]{AD-SDG}%
  \BibitemOpen
  \bibfield  {author} {\bibinfo {author} {\bibfnamefont {S.}~\bibnamefont
  {Bhattacharyya}}, \bibinfo {author} {\bibfnamefont {A.}~\bibnamefont {Das}},
  \ and\ \bibinfo {author} {\bibfnamefont {S.}~\bibnamefont {Dasgupta}},\
  }\href {\doibase 10.1103/PhysRevB.86.054410} {\bibfield  {journal} {\bibinfo
  {journal} {Phys. Rev. B}\ }\textbf {\bibinfo {volume} {86}},\ \bibinfo
  {pages} {054410} (\bibinfo {year} {2012})}\BibitemShut {NoStop}%
\bibitem [{\citenamefont {Das}\ and\ \citenamefont {Moessner}(2012)}]{AD-RM}%
  \BibitemOpen
  \bibfield  {author} {\bibinfo {author} {\bibfnamefont {A.}~\bibnamefont
  {Das}}\ and\ \bibinfo {author} {\bibfnamefont {R.}~\bibnamefont {Moessner}},\
  }\href@noop {} {\  (\bibinfo {year} {2012})},\ \Eprint
  {http://arxiv.org/abs/1208.0217} {arXiv:1208.0217} \BibitemShut {NoStop}%
\bibitem [{\citenamefont {Bukov}\ \emph {et~al.}(2014)\citenamefont {Bukov},
  \citenamefont {D'Alessio},\ and\ \citenamefont
  {Polkovnikov}}]{Anatoli-Periodic}%
  \BibitemOpen
  \bibfield  {author} {\bibinfo {author} {\bibfnamefont {M.}~\bibnamefont
  {Bukov}}, \bibinfo {author} {\bibfnamefont {L.}~\bibnamefont {D'Alessio}}, \
  and\ \bibinfo {author} {\bibfnamefont {A.}~\bibnamefont {Polkovnikov}},\
  }\href@noop {} {\  (\bibinfo {year} {2014})},\ \Eprint
  {http://arxiv.org/abs/1407.4803v2} {1407.4803v2} \BibitemShut {NoStop}%
\bibitem [{\citenamefont {Suzuki}\ \emph {et~al.}(2013)\citenamefont {Suzuki},
  \citenamefont {Inoue},\ and\ \citenamefont {Chakrabarti}}]{BKC-Book}%
  \BibitemOpen
  \bibfield  {author} {\bibinfo {author} {\bibfnamefont {S.}~\bibnamefont
  {Suzuki}}, \bibinfo {author} {\bibfnamefont {J.-i.}\ \bibnamefont {Inoue}}, \
  and\ \bibinfo {author} {\bibfnamefont {B.~K.}\ \bibnamefont {Chakrabarti}},\
  }\href@noop {} {\emph {\bibinfo {title} {Quantum Ising Phases and Transitions
  in Transverse Ising Models}}}\ (\bibinfo  {publisher} {Springer,
  Heidelberg},\ \bibinfo {year} {2013})\BibitemShut {NoStop}%
\bibitem [{\citenamefont {Dunlap}\ and\ \citenamefont {Kenkre}(1986)}]{Dunlap}%
  \BibitemOpen
  \bibfield  {author} {\bibinfo {author} {\bibfnamefont {A.~H.}\ \bibnamefont
  {Dunlap}}\ and\ \bibinfo {author} {\bibfnamefont {V.~M.}\ \bibnamefont
  {Kenkre}},\ }\href@noop {} {\bibfield  {journal} {\bibinfo  {journal} {Phys.
  Rev. B}\ }\textbf {\bibinfo {volume} {34}},\ \bibinfo {pages} {3625}
  (\bibinfo {year} {1986})}\BibitemShut {NoStop}%
\bibitem [{\citenamefont {Grossmann}\ \emph {et~al.}(1991)\citenamefont
  {Grossmann}, \citenamefont {Dittrich}, \citenamefont {Jung},\ and\
  \citenamefont {H{\"a}nggi}}]{CDT}%
  \BibitemOpen
  \bibfield  {author} {\bibinfo {author} {\bibfnamefont {F.}~\bibnamefont
  {Grossmann}}, \bibinfo {author} {\bibfnamefont {T.}~\bibnamefont {Dittrich}},
  \bibinfo {author} {\bibfnamefont {P.}~\bibnamefont {Jung}}, \ and\ \bibinfo
  {author} {\bibfnamefont {P.}~\bibnamefont {H{\"a}nggi}},\ }\href@noop {}
  {\bibfield  {journal} {\bibinfo  {journal} {Phys. Rev. Lett.}\ }\textbf
  {\bibinfo {volume} {67}},\ \bibinfo {pages} {516} (\bibinfo {year}
  {1991})}\BibitemShut {NoStop}%
\bibitem [{\citenamefont {Zhang~{\it et. al.}}(2008)}]{Ref2}%
  \BibitemOpen
  \bibfield  {author} {\bibinfo {author} {\bibfnamefont {J.}~\bibnamefont
  {Zhang~{\it et. al.}}},\ }\href
  {https://journals.aps.org/prl/abstract/10.1103/PhysRevLett.100.10050}
  {\bibfield  {journal} {\bibinfo  {journal} {Phys. Rev. Letts.}\ }\textbf
  {\bibinfo {volume} {100}},\ \bibinfo {pages} {100501} (\bibinfo {year}
  {2008})}\BibitemShut {NoStop}%
\bibitem [{\citenamefont {Zhang~{\it et. al.}}(2009)}]{Ref1}%
  \BibitemOpen
  \bibfield  {author} {\bibinfo {author} {\bibfnamefont {J.}~\bibnamefont
  {Zhang~{\it et. al.}}},\ }\href
  {https://journals.aps.org/pra/abstract/10.1103/PhysRevA.79.012305} {\bibfield
   {journal} {\bibinfo  {journal} {Phys. Rev. A}\ }\textbf {\bibinfo {volume}
  {79}},\ \bibinfo {pages} {012305} (\bibinfo {year} {2009})}\BibitemShut
  {NoStop}%
\bibitem [{\citenamefont {Lieb}\ \emph {et~al.}(1961)\citenamefont {Lieb},
  \citenamefont {Schultz},\ and\ \citenamefont {Mattis}}]{LSM}%
  \BibitemOpen
  \bibfield  {author} {\bibinfo {author} {\bibfnamefont {E.}~\bibnamefont
  {Lieb}}, \bibinfo {author} {\bibfnamefont {T.}~\bibnamefont {Schultz}}, \
  and\ \bibinfo {author} {\bibfnamefont {D.}~\bibnamefont {Mattis}},\
  }\href@noop {} {\bibfield  {journal} {\bibinfo  {journal} {Ann. Phys}\
  }\textbf {\bibinfo {volume} {16}},\ \bibinfo {pages} {407} (\bibinfo {year}
  {1961})}\BibitemShut {NoStop}%
\bibitem [{\citenamefont {Mattis}(2006)}]{Mattis}%
  \BibitemOpen
  \bibfield  {author} {\bibinfo {author} {\bibfnamefont {D.~C.}\ \bibnamefont
  {Mattis}},\ }\href@noop {} {\emph {\bibinfo {title} {The Theory of Magnetism
  Made Simple: An Introduction To Physical Concepts And To Some Useful
  Mathematical Methods}}}\ (\bibinfo  {publisher} {World Scientific.},\
  \bibinfo {year} {2006})\BibitemShut {NoStop}%
\bibitem [{\citenamefont {Damski}\ and\ \citenamefont {Rams}(2014)}]{Marek}%
  \BibitemOpen
  \bibfield  {author} {\bibinfo {author} {\bibfnamefont {B.}~\bibnamefont
  {Damski}}\ and\ \bibinfo {author} {\bibfnamefont {M.~M.}\ \bibnamefont
  {Rams}},\ }\href@noop {} {\bibfield  {journal} {\bibinfo  {journal} {J. Phys.
  A}\ }\textbf {\bibinfo {volume} {47}},\ \bibinfo {pages} {025303} (\bibinfo
  {year} {2014})}\BibitemShut {NoStop}%
\bibitem [{sup()}]{suppl}%
  \BibitemOpen
  \href@noop {} {}\bibinfo {note} {See the supplementary material.}\BibitemShut
  {Stop}%
\bibitem [{\citenamefont {Khaneja~\textit{et. al.}}(2005)}]{Khaneja}%
  \BibitemOpen
  \bibfield  {author} {\bibinfo {author} {\bibfnamefont {N.}~\bibnamefont
  {Khaneja~\textit{et. al.}}},\ }\href {\doibase 10.1016/j.jmr.2004.11.004}
  {\bibfield  {journal} {\bibinfo  {journal} {Journal of Magnetic Resonance}\
  }\textbf {\bibinfo {volume} {172}},\ \bibinfo {pages} {296} (\bibinfo {year}
  {2005})}\BibitemShut {NoStop}%
\bibitem [{\citenamefont {Kowalewski}\ and\ \citenamefont
  {Maler}(2006)}]{kowalewski2006nuclear}%
  \BibitemOpen
  \bibfield  {author} {\bibinfo {author} {\bibfnamefont {J.}~\bibnamefont
  {Kowalewski}}\ and\ \bibinfo {author} {\bibfnamefont {L.}~\bibnamefont
  {Maler}},\ }\href@noop {} {\emph {\bibinfo {title} {Nuclear spin relaxation
  in liquids: theory, experiments, and applications}}},\ Vol.~\bibinfo {volume}
  {2}\ (\bibinfo  {publisher} {Taylor \& Francis},\ \bibinfo {year}
  {2006})\BibitemShut {NoStop}%
\bibitem [{\citenamefont {Mukherjee}\ \emph {et~al.}(2008)\citenamefont
  {Mukherjee}, \citenamefont {Dutta},\ and\ \citenamefont {Sen}}]{Amit-deco}%
  \BibitemOpen
  \bibfield  {author} {\bibinfo {author} {\bibfnamefont {V.}~\bibnamefont
  {Mukherjee}}, \bibinfo {author} {\bibfnamefont {A.}~\bibnamefont {Dutta}}, \
  and\ \bibinfo {author} {\bibfnamefont {D.}~\bibnamefont {Sen}},\ }\href
  {\doibase 10.1103/PhysRevB.77.214427} {\bibfield  {journal} {\bibinfo
  {journal} {Phys. Rev. B}\ }\textbf {\bibinfo {volume} {77}},\ \bibinfo
  {pages} {214427} (\bibinfo {year} {2008})}\BibitemShut {NoStop}%
\end{thebibliography}%


\begin{thebibliography}{99}   



%
%
%
%
%
%
\bibitem{zurek1}
W. H. Zurek, Rev. Mod. Phys. {\bf 75}, 3 (2003).

\bibitem{cavanagh}
J. Cavanagh, W. J. Fairbrother, A. G. Palmer III and N. J. Skelton,
\textit{Protein NMR spectroscotpy: Principles and Practice}
(Academic Press, 1995).


\bibitem{T1rho}
J. Kowalewski and L. Maler, 
\textit{Nuclear spin relaxation in liquids: theory, experiments, and applications}, Vol. 2

\end{thebibliography}

\section{Supplementary Material}
\subsection{Thermal Equilibrium State in NMR}
The NMR density matrix of an ensemble of a system of $ n $ spin-$ 1/2 $ homonuclear Nucleus (i.e. having same gyromagnetic ratio $ \gamma $)
in thermal equilibrium at an ambient temperature $ T $ and inside a uniform magnetic fiels $ B_0 $ in $ z $-direction is \cite{cavanagh}
\begin{equation}
\rho = \frac{1}{Z}\exp(-{\cal H}/kT).
\end{equation}
Here
\begin{equation}
{\cal H} = \hbar\gamma B_0\sum\limits_{j=1}^{n} I_j^z,
\end{equation}
is the Hamiltonian , $  Z = \mathrm{Tr[\exp(-{\cal H}/kT)]} = 2^n $ is the partition function and $ k $ is the Boltzman Constant.
Under the high-temperature and high-field approximation ($ kT\gg \Delta E = \hbar \gamma B_0 $, the energy gap), the above form can be expanded to 

\begin{eqnarray}
\rho &\approx& \frac{1}{2^n} \mathbbm{1} - {\cal H}/kT \nonumber \\
&=& \frac{1}{2^n} \mathbbm{1} + \epsilon \rho_{\Delta},
\end{eqnarray}
where $ \epsilon = \gamma \hbar B_0 / kT$, and $ \rho_{\Delta} = \sum\limits_{j=1}^{n} I_j^z$ is the traceless deviation part.

Since the Identity part does not transform under any Unitary transformation and does not give any signal in NMR, we ignore it. So in all practical
calculations, we simply take the trace-less part $ \rho_{\Delta} = \sum\limits_{j=1}^{n} I_j^z $ or $ \rho_{\Delta} = \sum\limits_{j=1}^{n} \sigma_j^z $ (absorbing $ 1/2 $ in
$ \epsilon $) as thermal state NMR density matrix.

\subsection{Signal in NMR}
The real part of the signal (i.e. the magnetization along the $ x $ direction), $ S(t) $ in NMR is \cite{cavanagh}
\begin{equation}
S(t) \propto \mathrm{Tr}[\rho_t D],
\end{equation}
where $ \rho_t = U_t \rho_{\Delta} U_t^{\dagger}$ is the instantaneous density matrix after the Unitary transformation given by $ U_t $, and the detection 
operator $ D=\sum\limits_{j=1}^{n}I_j^z $. To get rid of the proportionality constant, we normalize the signal with respect to the
signal obtained after applying a global $ (\pi/2)_y $ pulse on the thermal equilibrium state (i.e. $ \rho_t = \sum\limits_{j=1}^{n} \sigma_j^x $).
\begin{equation}
S_{norm}(t) = \frac{\mathrm{Tr}[\rho_t D]}{\mathrm{Tr}[\sum\limits_{j=1}^{n} \sigma_j^x D]} = \frac{1}{12}\mathrm{Tr}[\rho_t D]  \mathrm{\ \ \ \  \ \ for\  } n = 3.
\end{equation}

\subsection{Decoherence and Inverse Decay}
\begin{figure}[h]
\centering
\includegraphics[width=0.95\linewidth]{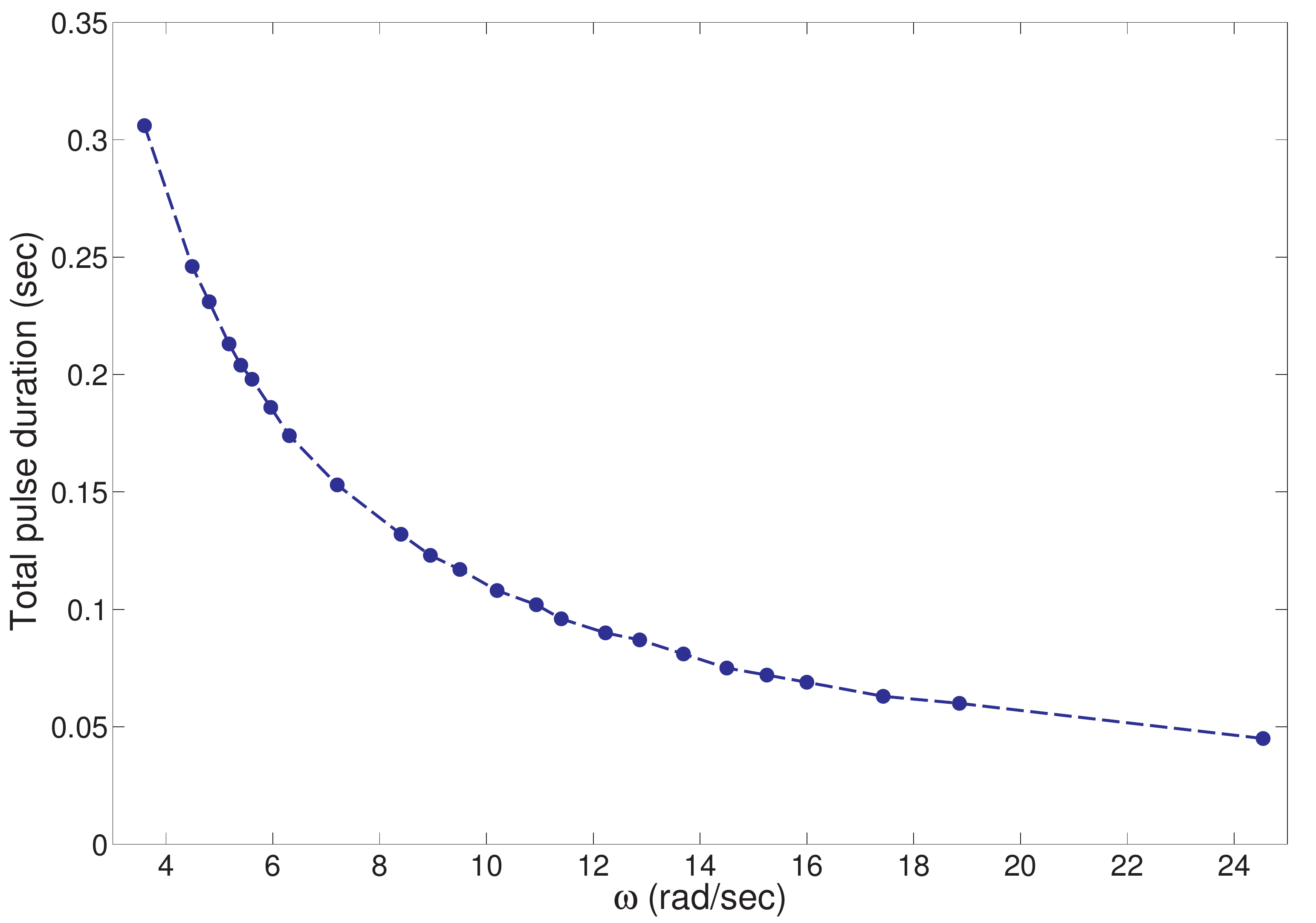}
\caption{(Color online)
The total experimental pulse time for various $\omega$ values.     
}
\label{tot_t}
\end{figure}

Quantum systems constantly interact with their neighboring environment. This makes the quantum systems liable to irreversible loss of coherences leading to decoherence \cite{zurek1}. The single spin decoherence times are characterized by transverse relaxation time constants ($T_2$). Their values for $F_1$, $F_2$ and $F_3$ are measured by Hahn echo \cite{cavanagh} technique and are found to be $2.8$, $3.1$, and $3.3$ s respectively. Fig. \ref{tot_t} gives the experimental pulse time duration for a range of chosen $\omega$ values. We see that lower $\omega$ values require more time since $\tau=2\pi/\omega$ and this indicates the faster decay of magnetization for very low $\omega$ values. In the presence of RF modulation, as in our case, the transverse magnetization decays with an average decay constant $T_{1\rho}$ \cite{T1rho}. For multispin coherences, $T_{1\rho}$ can be shorter than single spin $T_2$ values. Moreover, lower fidelity pulses and RF inhomogeneity can result in faster magnetization decay which accounts for a overall relaxation time of $T_d$. Here, we explain the model for the decay of magnetization and its correction using inverse decay method in step 1 and step 2 respectively. 
 
{\it step 1 - Determining $T_d$}: 

The decay model for the evolution of transverse magnetization ($m^x(t)$) 
due to the decay of amplitude and the overall envelope under the relaxation time $T_d$ is given by
\begin{equation}
m^x(t) = \alpha+[\beta+\gamma \cos(ct)]e^{-t/T_d}
\label{m1}
\end{equation}
where $\alpha$, $\beta$, $\gamma$, $c$ and $T_d$ are the fitting parameters. 
The model is motivated by the analytical form of $m^x(t)$ (Eq. 11 from the main manuscript), 
and the fact that a quantity without any explicit time-independence
suffers from standard exponential decay due to RF inhomogeneity and small pulse errors. 
In general, these fitting parameters can be measured by fitting the experimental 
$m^x(t)$ to Eq. 6. All the experiments were performed with RF pulses with fidelities greater than 
or equal to 0.99. These small imperfections in the making of 
pulses along with the RF inhomogeneity resulted in decay of the 
signal $S(t)$ ($\propto m^x(t)$) that was reflected in lower $T_d$ values. 
Further, the decay of $m^x(t)$ decreases the $Q$ value since $Q$ is the long 
time average of $m^x(t)$. This explains the overall downward shifting of the 
raw experimental $Q$ data (Fig. 4 in the main manuscript).
\begin{figure*}
\centering
\includegraphics[width=0.7\linewidth]{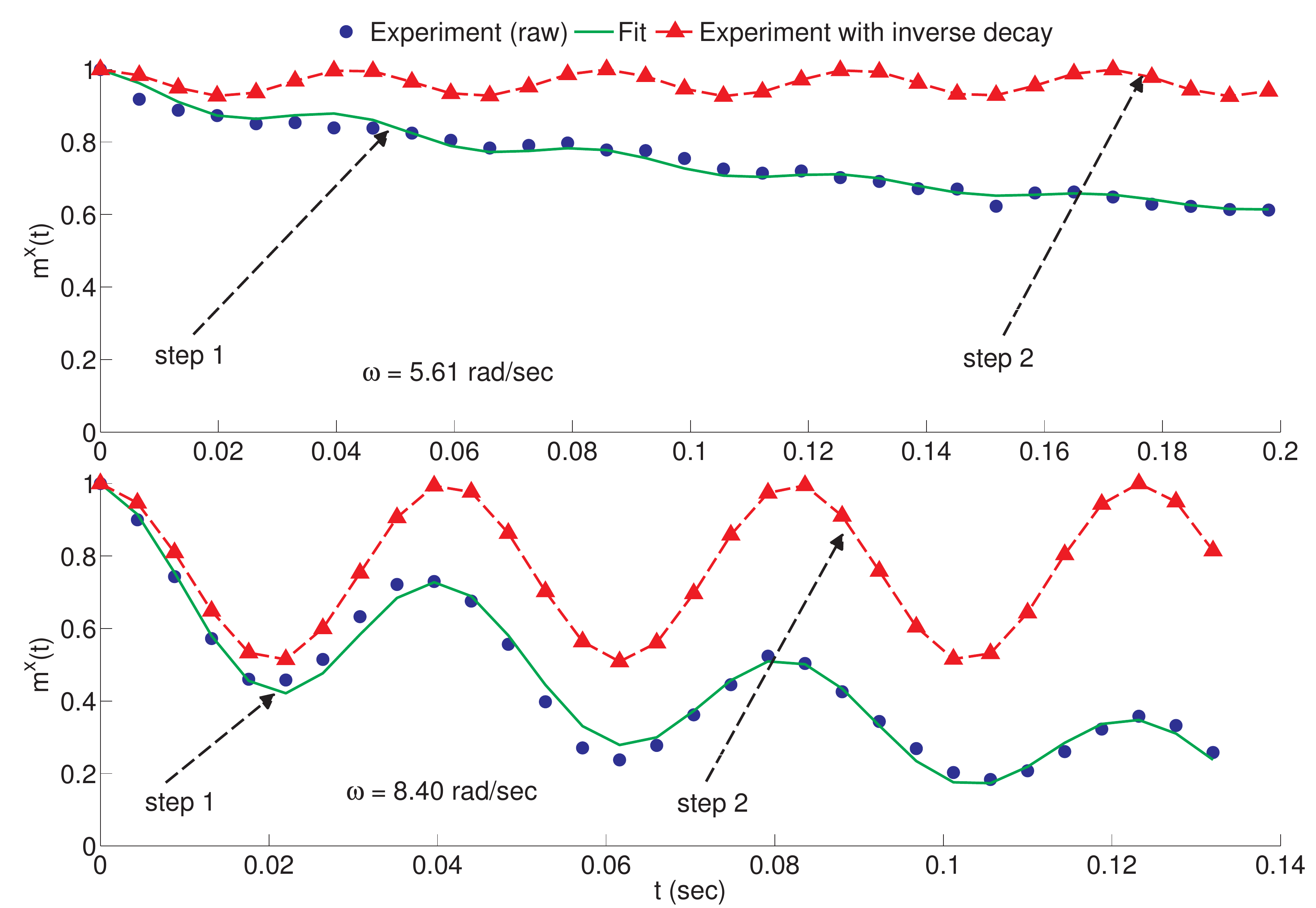}
\caption{(Color online)
An example illustrating the method to obtain $m^x(t)$ with inverse decay correction for $\omega=5.61$ rad/sec (freezing) and $\omega = 8.40$ rad/sec (non-freezing). Here $m^x(0)=1$, $h_0 = 5\pi$ and ${\cal J} = h_0/20$ in both the cases. Step 1 involves determining $\alpha$, $\beta$, $\gamma$, $c$ and $T_d$ while step 2 gives $m^x(t)$ that takes the fitting parameters as inputs with $T_d \rightarrow \infty$. 
}
\label{inv}
\end{figure*}

{\it step 2 - Inverse decay}: 

This method illustrates a data processing technique that gives the information about $m^x(t)$ in the absence of any relaxation phenomenon. This is an ideal situation where $T_d\rightarrow \infty$. 
Thus, in the absence of decay, Eq. 6 becomes
\begin{equation}
m^x(t) = \alpha+[\beta+\gamma \cos(ct)]
\label{invd}
\end{equation}
Fig. \ref{inv} shows an explicit demonstration for obtaining $m^x(t)$ using inverse decay method.
The table below shows $Q$ (for raw experiments), $T_{d}$, and $Q$ with inverse decay for various $\omega$ values. 
\begin{center}

\begin{tabular}{|c|c|c|c|}

\hline
$\omega$ (rad/sec)  & $Q$ (raw)& $T_{d}$ (s) & $Q$ with inverse decay  \\
\hline
3.59 &   0.54  &  0.06 &   0.95\\
4.49 &   0.46  &  0.09  &  0.83\\
4.81  &  0.69  &  0.28 &   0.88\\
5.18  &  0.76 &   0.39 &   0.94\\
5.40 &   0.84 &   0.08 &   0.95\\
5.61  &  0.75 &   0.17   & 0.96\\
5.96  &  0.88 &   0.07  &  0.96\\
6.31 &   0.68 &   0.22  &  0.92\\
7.21 &   0.68  &  0.17  &  0.76\\
8.40  &  0.53  &  0.15  &  0.79\\
8.95  &  0.68 &   0.49  &  0.79\\
9.50 &   0.72  &  0.14  &  0.80\\
10.20  &  0.76 &   0.16&    0.83\\
10.93 &   0.84  & 0.09   & 0.89\\
11.40 &   0.85  &  0.07 &   0.92\\
12.23 &   0.89 &   0.03 &   0.95\\
12.87 &   0.89 &   0.02 &   0.94\\
13.69 &   0.88 &   0.05 &   0.95\\
14.50 &   0.85 &   0.05 &   0.90\\
15.25 &   0.81 &   0.08 &   0.87\\
16.00  &  0.77 &   0.11 &   0.84\\
17.43 &   0.71  &  0.14 &   0.77\\
18.85  &  0.63  &  0.13 &   0.74\\
24.54 & 0.49  &  0.11  &  0.65\\
\hline
\end{tabular}
\end{center}

\subsection{Error Bars}
The error bars introduced in Fig. 4 of the main manuscript correspond to the RMS value of the errors with respect to magenization evolution for various $\omega$ values. The way we calculated RMS error bars are as below:
\begin{enumerate}
\item Let, for example, the raw experimental values of transverse magnetization be $M^{x}_{raw}(t)$.
\item Let $M^{x}_{fit}(t)$ be the magnetization values after fitting $M^{x}_{raw}(t)$ with Eq. 6.
\item The difference between $M^{x}_{raw}(t)$ and $M^{x}_{fit}(t)$ give the errors. We calculated the RMS value of these errors and these RMS errors are reflected in Q values since $Q = \frac{1}{N+1}\sum\limits_{n=0}^{N}M^x(n\tau)$ (defined in MS).

\end{enumerate}

\end{document}